\documentclass[aps,prb,twocolumn,amsmath,amssymb,superscriptaddress]{revtex4-1}

\usepackage{graphicx}  
\usepackage{physics}   
\usepackage{xcolor}    
\usepackage{hyperref}  
\hypersetup{
    colorlinks=true,
    linkcolor=blue,
    citecolor=blue,
    filecolor=magenta,
    urlcolor=blue,
    pdftitle={Dynamically protected erasure qubit via low-frequency charge driving},
    }
\setcitestyle{super,sort&compress}   

\usepackage[resetlabels]{multibib}
\makeatletter
\let\supp@bibcite\bibcite
\def\bibcite#1{\supp@bibcite{#1@supp}}
\newcites{supp}{Supplementary References}
\let\bibcite\supp@bibcite
\AtBeginDocument{\let@environment{thebibliography}{rtx@thebibliography}}
\newcommand{\suppcitesbegin}{%
 \def\@extra@b@citeb{@supp}%
 \expandafter\let\expandafter\mb@main@cite\csname cite \endcsname
 \expandafter\def\csname cite \endcsname{%
  \let\@citex\mb@@citex
  \let\@newciteauxhandle\@auxoutsupp
  \mb@main@cite}%
 \let\supp@citenumfont\citenumfont
 \let\supp@bibnumfmt\bibnumfmt
 \renewcommand{\citenumfont}[1]{\supp@citenumfont{S##1}}%
 \renewcommand{\bibnumfmt}[1]{\supp@bibnumfmt{S##1}}%
}
\newcommand{\suppbibliography}[1]{%
 \begingroup
  \let\label\@gobble
  \let\pre@bibdata\@empty
  \let\@FMN@list\@empty
  \let\bibsection\rtx@bibsection
  \bibliographystylesupp{\@bibstyle}%
  \bibliographysupp{#1}%
 \endgroup
}
\makeatother

\begin{document}

\title{Dynamically protected erasure qubit via low-frequency charge driving}

\author{Ahmed Hajr}
\thanks{Correspondence to \url{hajr@berkeley.edu}}
\affiliation{Quantum Nanoelectronics Laboratory, Department of Physics, University of California at Berkeley, Berkeley, California 94720, USA}

\author{Ke Wang}
\affiliation{Quantum Nanoelectronics Laboratory, Department of Physics, University of California at Berkeley, Berkeley, California 94720, USA}
\affiliation{Computational Research Division, Lawrence Berkeley National Laboratory, Berkeley, California 94720, USA}

\author{Bingcheng Qing}
\affiliation{Quantum Nanoelectronics Laboratory, Department of Physics, University of California at Berkeley, Berkeley, California 94720, USA}

\author{Akel Hashim}
\affiliation{Quantum Nanoelectronics Laboratory, Department of Physics, University of California at Berkeley, Berkeley, California 94720, USA}
\affiliation{Computational Research Division, Lawrence Berkeley National Laboratory, Berkeley, California 94720, USA}

\author{Bibek Bhandari}
\affiliation{Institute for Quantum Studies, Chapman University, Orange, CA, USA}

\author{Dominic Briseño-Colunga}
\affiliation{Institute for Quantum Studies, Chapman University, Orange, CA, USA}

\author{Justin Dressel}
\affiliation{Institute for Quantum Studies, Chapman University, Orange, CA, USA}

\author{Andrew N. Jordan}
\affiliation{Department of Physics and Astronomy, University of Rochester, Rochester, NY, USA}
\affiliation{Institute for Quantum Studies, Chapman University, Orange, CA, USA}
\affiliation{The Kennedy Chair in Physics, Chapman University, Orange, CA 92866, USA}

\author{Irfan Siddiqi}
\affiliation{Quantum Nanoelectronics Laboratory, Department of Physics, University of California at Berkeley, Berkeley, California 94720, USA}
\affiliation{Computational Research Division, Lawrence Berkeley National Laboratory, Berkeley, California 94720, USA}

\begin{abstract}
    Dynamical protection via strong driving can enable resilient quantum processing on imperfect physical hardware. However, the practical utility of such schemes is frequently limited by parasitic processes such as drive-induced dephasing and leakage. Here, we demonstrate that sub-GHz charge driving of superconducting Kerr oscillators (KOs) in the transmon regime circumvents this trade-off, simultaneously activating fast parametric interactions and protecting the encoded qubit from low-frequency noise. The key is the frequency dependence of the charge sensitivity: using the AC Stark shift as a probe, we find that the KO frequency sensitivity to a charge drive grows quadratically with both drive frequency and amplitude, making the oscillator weakly sensitive to $1/f$ charge noise yet strongly coupled to drives near 1~GHz. Exploiting this, we first demonstrate cooling and reset of the KO in 82~ns, to a residual population below $0.7\%$—lower than its $2.5\%$ steady-state thermal population. Next, using two KOs, we demonstrate logical control and dynamical protection of a dual-rail qubit with nearly fourfold erasure bias. Finally, using a single end-of-circuit erasure check, we achieve an error per Clifford of $5.6\times10^{-4}$, which falls to $1.5\times10^{-4}$ after post-selection, with 25~ns gates.
\end{abstract}

\maketitle

\vspace{1em}
\noindent\textbf{Introduction}

\noindent A primary challenge in experimental quantum computing is engineering noise-resilient systems while minimizing the hardware overhead required for universal control and error correction \cite{divincenzo2000physical,Martinis_QEC,preskill_nisq}. Superconducting circuits, as a platform, typically leverage the charge sensitivity of Josephson junction-based oscillators to implement single- and two-qubit gates \cite{martinis1985energy,wallraff2004strong,blais2004cavityqed,majer2007coupling}. While high-fidelity operations can be activated via resonant or near-resonant microwave drives \cite{Rigetti_CR,CR_ibm, mitchell2021hardware, nguyen2024empowering}, these systems remain susceptible to low-frequency $1/f$ noise, a ubiquitous feature of solid-state environments \cite{1_f_noise,1_f_flux_noise,Muller_TLS}. To suppress the sensitivity of the superconducting Kerr oscillator (KO) to low-frequency charge noise, it is typically operated in the transmon regime, where the oscillator anharmonicity is a small fraction of the oscillator frequency \cite{koch2007charge,schreier2008suppressing}. This architecture proved very promising, with coherence times on the order of hundreds of microseconds \cite{wang2022transmonnpj,place2021tantalum}, and multi-qubit processors capable of demonstrating quantum advantage \cite{Quantum_supremacy_nature,Quantum_supremacy_PRL} and quantum error correction \cite{QEC_1,QEC_2,QEC_3,QEC_4}. Nonetheless, unlike flux driving \cite{Hime2006_Science,Flux_noise_spectroscopy_NatPhys,Driven_dynamics_and_rotary_echo_PRL,Rotating_frame_relaxation_NatComm,Fast_Tunable_Coupling_PRL,Tunable_Coupling_Scheme_PhysRevApplied,STS_PRXQuantum,LINC_PRXQuantum}, a comprehensive study of the sensitivity of the superconducting KO to low-frequency charge driving and its utility in wave mixing processes is still lacking.

In this context, it is important to distinguish between two effects: the charge sensitivity due to the periodicity of the cosine potential in the Josephson junction \cite{josephson1962_PhysLett,Devoret2021_JJ} and the charge sensitivity to wave mixing processes in the superconducting KO. The first effect manifests as a periodic charge dispersion even at DC voltages and can be exponentially suppressed by reducing the zero-point fluctuations in the phase operator \cite{koch2007charge} or extending the energy potential with multiple Josephson junctions \cite{Manucharyan_fluxonium}  (effectively increasing the oscillator frequency to anharmonicity ratio). The second effect depends on the frequency of the charge drive and scales proportionally with the nonlinearity of the superconducting KO. We focus on the latter and demonstrate its utility in performing a variety of high-fidelity quantum operations with sub-GHz charge drives.

In this work, we use a planar (2D) superconducting circuit with two KOs that features strong KO--KO coupling, strong charge coupling to the drive lines, and low drive crosstalk. This architecture enables us to perform two key studies: first, we examine the AC Stark shift to determine the KO frequency sensitivity to low-frequency charge driving \cite{MIST_prx}; second, we leverage the third-order nonlinearity of the KOs to execute efficient beam-splitter interactions between different circuit modes \cite{frattini2017snail,Beam_Splitter_PRXQuantum}. The Stark-shift measurements reveal that the sensitivity of the superconducting KO frequency to low-frequency charge driving increases quadratically with the drive frequency. While this scaling is favorable because it counters the $1/f$ spectrum of the noise, it also imposes a design challenge: to achieve efficient wave mixing, we must maintain strong coupling ($g/2\pi \sim 50$--$200$ MHz) and detuning ($\Delta/2\pi \sim 0.5$--$2$ GHz) between the circuit modes ($\omega/2\pi \sim 5$--$8$ GHz) with suppressed microwave cross-talk to the drive lines. 

\begin{figure*}[tph!]
    \includegraphics[width=\textwidth]{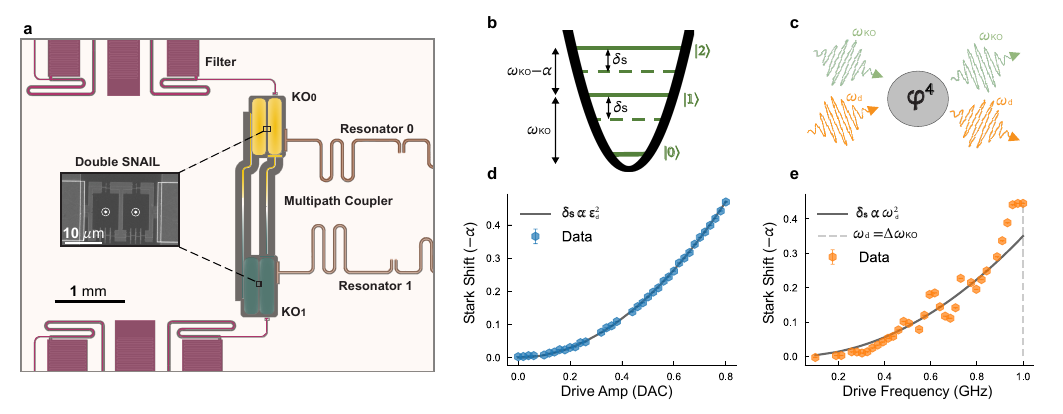}
    \caption{\label{fig1}\textbf{Superconducting KO chip and charge sensitivity measurements}. 
    \textbf{a,} Experimental schematic for the 2D chip. Each KO is a double SNAIL capacitively shunted in a floating design (yellow and teal). The readout is performed using the dispersive interaction between the KOs and the readout resonators with dedicated Purcell filters (orange). The coupling between the KOs is implemented in a multi-path design to minimize the crosstalk to the microwave control lines. The band-stop filter (dark red) suppresses the Purcell decay of the KOs while enabling strong coupling off-resonance. The design parameters are discussed in more detail in the Supplementary Note \ref{note:device}. \textbf{b,} The effect of a strong off-resonance drive on the KO with anharmonicity $-\alpha$ can be estimated from the Stark shift $\delta_s$ on the energy levels. \textbf{c,} The four-wave mixing process underlying the AC Stark shift. \textbf{d,} The AC Stark shift as a function of the drive amplitude in digital-to-analog converter (DAC) units while fixing the drive frequency to 0.85 GHz. The quadratic scaling results from two drive photons participating in the process. \textbf{e,} The AC Stark shift as a function of the drive frequency when the drive amplitude is set to 0.6 (DAC). The quadratic scaling results from the destructive interference between the rotating and counter-rotating terms at low frequencies.}
\end{figure*}

Using this device, we first demonstrate that low-frequency three-wave mixing can reset the KO to less than $0.7\%$ population in 82 ns, which is smaller than the steady-state thermal population of $2.5\%$. Next, we characterize the coherent beam-splitter interaction between the two KOs and use it to dynamically protect the system from low-frequency noise. Finally, we encode a dual-rail qubit \cite{teoh2023dualrail,kubica2023erasure,levine2024dualrail,Mehta2026Nature} in the two KOs and demonstrate its fourfold erasure bias, achieving a single-qubit error per Clifford of $5.6\times 10^{-4}$ ($1.5\times 10^{-4}$ after erasure post-selection) with 25-ns gate time.

\vspace{0.5em}

\noindent\textbf{KO low-frequency charge sensitivity}

\noindent The superconducting chip used to study the charge sensitivity of the KO in this work is shown in Fig.~\ref{fig1}\textbf{a}. It contains two KOs made of capacitively shunted double Superconducting Nonlinear Asymmetric Inductive eLements (SNAILs) \cite{frattini2017snail}. The utility of the SNAIL in this work is to provide third-order and fourth-order nonlinearities while keeping the DC charge dispersion of the KO negligible (supplemental note \ref{note:snail_nonlinearity}). The relatively small fourth-order nonlinearity is used to study the charge sensitivity using the AC Stark shift, while the third-order nonlinearity is used to perform the beam-splitter interactions between the different modes of the circuit, as discussed in the next sections. The microwave control lines are capacitively coupled to the KOs with band-stop filters to suppress the Purcell decay \cite{hajr2024kerrcat2d}. The readout resonators enable us to perform dispersive, single-shot measurements of the state of the KOs \cite{wallraff2004strong}. For more details on the device parameters, we refer to the Supplemental Note \ref{note:device}.

To study the charge sensitivity of the KO, we perform two-tone spectroscopy on the KO--Resonator system in the presence of the charge drive to observe the AC Stark shift ($\delta_s$) as depicted in Fig.~\ref{fig1}\textbf{b}. The KO frequency in this part was tuned to $\omega_{\text{KO}}/2\pi=5.399$ GHz with anharmonicity of $-\alpha/2\pi=-5.3$ MHz. In this range, the charge dispersion of the KO is negligible, and the dominant source of the Stark shift is the four-wave mixing process mediated by Kerr nonlinearity of the oscillator. This process is shown in Fig.~\ref{fig1}\textbf{c}, where a photon from the drive and the KO are simultaneously destroyed and recreated, resulting in a shift in the KO frequency. At low-frequency the AC Stark shift can be expressed in terms of the anharmonicity of the KO and the displacement caused by the charge drive as (Supplemental Note \ref{note:wave_mixing}):

\begin{equation}
   \hat{\mathcal{H}}_s= -2 \alpha |\xi_d|^2  \hat{a}^\dagger \hat{a} \approx -8 \alpha \left(\frac{|\epsilon_d| \omega_d}{\omega^2_{\text{KO}}}  \right)^2 \hat{a}^\dagger \hat{a}
\end{equation}

where $\epsilon_d$ is the drive amplitude and $\omega_d$ is the drive frequency. This formula shows that the AC Stark shift is quadratic in both drive parameters, which is advantageous over standard charge dispersion or flux modulation because it suppresses the effects of near-DC noise while enabling efficient interaction with near-GHz drives. 

Fig.~\ref{fig1}\textbf{d} shows the AC Stark shift as a function of the drive amplitude when the drive frequency is set to 0.85 GHz. The quadratic scaling is expected because this process involves two photons from the drive. However, the surprising result is the scale of this Stark shift, given the drive parameters. This issue is discussed in the Supplemental Note \ref{note:wave_mixing}.

The dependence of the AC Stark shift on the drive frequency is shown in Fig.~\ref{fig1}\textbf{e}. The quadratic scaling of the Stark shift (i.e., linear scaling in the displacement) results from the destructive interference between the rotating and counter-rotating terms at low frequencies, with some ripples due to higher-order resonances with the circuit modes. The practical benefit of this scaling is the suppression of low-frequency noise, but the design constraint is the need for large detunings between the circuit modes to enable strong interactions with charge driving (i.e., large drive-induced displacement).

\begin{figure}[tp]
    \includegraphics[]{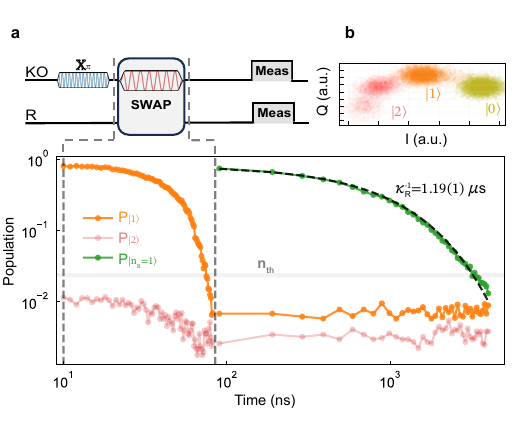}
    \caption{\label{fig2}\textbf{KO parametric reset and cooling}. 
    \textbf{a,} Pulse sequence for the parametric reset and cooling of the KO and the corresponding time evolution. The region between the dashed lines shows the coherent conversion of an excitation from the KO into an excitation in the lossy resonator, completed after 82 ns. The free evolution afterwards shows the exponential decay of the oscillator and the cooling of the KO below its thermal population. \textbf{b,} The demodulated readout signal in the IQ plane showing the different blobs associated with the ground state $\ket{0}$, the first excited state $\ket{1}$, and the second excited state $\ket{2}$. The effect of MIST is most visible of the second excited state.}
\end{figure}
\vspace{0.5em}
\noindent\textbf{KO reset and cooling}

\noindent The first demonstration of the utility of low-frequency charge driving is the parametric reset and cooling of the KO. Both processes are crucial not only for initializing the quantum system in a known state but also for mitigating leakage errors and performing reliable quantum error correction \cite{Miao2023Overcoming,leakage_reduction_PRL}. This process utilizes the third-order nonlinearity of the SNAIL to convert an excitation from the KO into an excitation in the lossy resonator by driving at the frequency difference between the modes (i.e., $\omega_d=\omega_R-\omega_{\text{KO}}$). Unlike most qubit reset schemes that rely on the resonator decay rate being much larger than the interaction strength (i.e., overdamped evolution) \cite{magnard2018allmicrowavereset,sunada2022fastreadoutreset,zhou2021rapidparametricreset}, we push this process to the coherent swap limit, where the interaction strength exceeds the resonator decay rate. The coherent swap nature of this process makes it ideal for mitigating leakage without introducing excess errors in the code space with larger anharmonicity devices. 

The protocol for this process is shown in Fig.~\ref{fig2}\textbf{a}. We first apply an $X(\pi)$ pulse to prepare the KO in the excited state, followed by a beam-splitter drive of variable duration to implement the exchange of excitations. The region between the two vertical dashed lines shows the sinusoidal population transfer out of the KO to achieve near-ideal exchange with the resonator after 82 ns. After the exchange, the free evolution of the system consists of pure decay from the resonator into the 50 $\Omega$ transmission line on a time scale of $\kappa_R^{-1}=1.1 \, \mu$s, with no observable change in the KO population. The KO population is measured using the dispersive shift in a readout signal at the resonator frequency, and the resonator population is measured by swapping the excitation back to the KO and performing the same dispersive readout. 

The estimated KO population at the end of this protocol is bounded above by $ 0.7\%$ inferred from the classified readout signal in the IQ plane shown in Fig.~\ref{fig2}\textbf{b}. This value is typical for measurement-induced state transition (MIST) \cite{MIST_prx} which is more visible on the second excited state (more details in the Supplementary Note \ref{note:dressed_states}). Nonetheless, this scheme leads to a lower population in the KO compared to the thermal population of $n_{\text{th}}=2.5 \%$ measured after an idle time much longer than the KO lifetime. This trend of active reset leading to a lower residual population was also observed in similar reset protocols of superconducting qubits \cite{zhou2021rapidparametricreset}.

\begin{figure*}[htbp!]
    \includegraphics[]{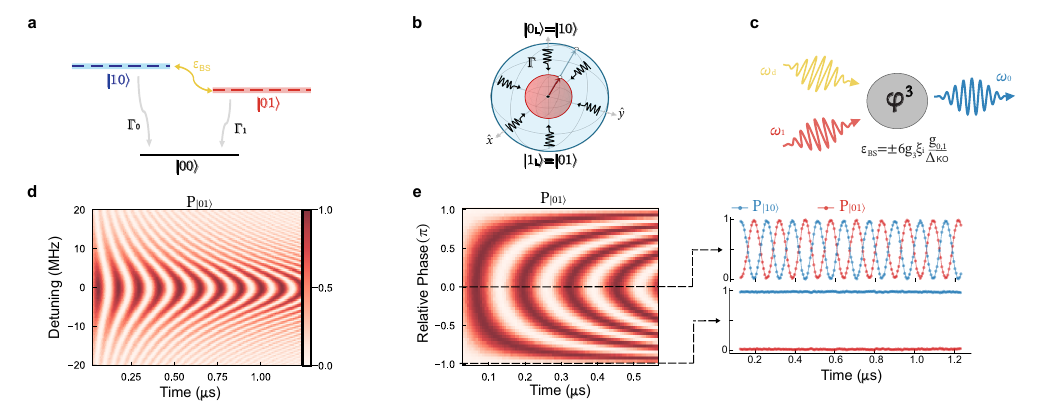}
    \caption{\label{fig3}\textbf{Dual-rail encoding and control}. 
    \textbf{a,} Simplified energy diagram of the KOs used in the dual-rail encoding. The beam-splitter interaction connects the single-excitation manifold, acting as a logical Rabi drive, while the action of single-photon loss maps the excited states to the ground state of the system. \textbf{b,} The Bloch sphere of the dual-rail qubit showing the logical code words $\ket{0_L}=\ket{10}$ and $\ket{1_L}=\ket{01}$ and the action of single-photon loss as shrinking in the Bloch sphere. \textbf{c,} The three-wave mixing process associated with the beam-splitter interaction mediated by each KO with the corresponding hybridization $\pm g_{0,1}/\Delta_{\text{KO}}$ and displacement $\xi_i$. \textbf{d,} The Rabi oscillations in the population $P_{01}$ of the dual-rail qubit as a function of the drive detuning from $\Delta_{\text{KO}}/2\pi$. The data are post-selected on the absence of single-photon loss at the end of each run to improve the SNR. \textbf{e,} The Rabi oscillations of the dual-rail qubit as a function of the relative phase between the contributions of KOs to $\epsilon_{\text{BS}}$. Two slices are shown when the contributions add up (top) or perfectly cancel (bottom).}
\end{figure*}
\vspace{0.5em}
\noindent\textbf{Dual-rail qubit encoding and control}

\noindent The second demonstration of the utility of low-frequency charge driving is the logical control and dynamical protection of the dual-rail qubit using two KOs. While this scheme relies on the same three-wave mixing process introduced in the previous section, utilizing two modes with relatively long lifetimes allows us to investigate dynamical protection arising from spin-locking to the charge drive. In this section, we review the dual-rail encoding and its realization in our system; we then explore spin-locking and dynamical protection in greater detail in the following sections.

The dual-rail encoding defines a qubit using single-photon excitations in two distinct modes. As shown in Fig.~\ref{fig3}\textbf{a}, this structure forms an inverted lambda system in which single-photon loss maps the code space to the ground state.  Consequently, this encoding satisfies the Knill-Laflamme criteria for quantum error detection of single-photon loss \cite{knill1997theory}, making it particularly advantageous in architectures where single-photon loss is the dominant error channel. In that case, the erasure bias can be leveraged for more efficient decoding when the qubit is embedded in a larger code space \cite{gu2025optimizing,sahay2023highthreshold}. A popular realization of this encoding is superconducting 3D cavities, which exhibit very little pure dephasing \cite{teoh2023dualrail,chou2024superconducting,Mehta2026Nature}. Also, more recently, encoding the dual-rail qubit with strongly hybridized modes was proposed \cite{kubica2023erasure} and demonstrated \cite{levine2024dualrail} as a robust alternative to suppress low-frequency noise in general. Our system operates on a similar principle, leveraging spin-locking to a charge drive to dynamically protect the qubit against low-frequency noise to establish the erasure bias.

Adopting the dual-rail encoding yields the logical Bloch sphere depicted in Fig.~\ref{fig3}\textbf{b}, spanned by the logical states $\ket{0_L}=\ket{10}$ and $\ket{1_L}=\ket{01}$. In this setup, single-photon loss uniformly shrinks the Bloch sphere without inducing transitions within the code space, provided the KOs have the same decay rate ($\Gamma_0=\Gamma_1$) \cite{PhysRevX.6.031006} or exhibit strong inter-mode coupling \cite{Naghiloo2019NaturePhysics}. This coupling can be static \cite{levine2024dualrail} or dynamic, as in our case. Moreover, in this encoding, the qubit is insensitive to the common frequency noise on the KOs. 

Universal single-qubit gates in this encoding are implemented using the beam-splitter interaction between the KOs as shown in Fig.~\ref{fig3}\textbf{c} with virtual $Z$ rotations \cite{virtual_Z_gate_PhysRevA}. Since both of the KOs are capable of mediating this beam-splitter interaction, we drive them simultaneously at their frequency difference $\omega_d=\Delta_{\text{KO}}=\omega_0-\omega_1$, which activates the interaction term in the Hamiltonian:

\begin{equation}
    \hat{\mathcal{H}}_\text{BS}=\epsilon_{\text{BS}}\hat{a}_0^{\dagger}\hat{a}_1+ \epsilon^*_{\text{BS}} \hat{a}_0 \hat{a}_1^{\dagger}
    \label{eq:BS_Hamiltonian}
\end{equation}

The interaction strength $\epsilon_{\text{BS}}$ to leading order is proportional to the displacement caused by each charge drive according to:

\begin{equation}
    \epsilon_{\text{BS}}=6g_{3,0}\xi_0 \frac{g_{0,1}}{\Delta_{\text{KO}}}-6g_{3,1}\xi_1 \frac{g_{0,1}}{\Delta_{\text{KO}}}
    \label{eq:BS_strength}
\end{equation}

The opposite signs of the two contributions originate from the dressed-state
hybridization of the modes. So, driving the two KOs with opposite displacements
(i.e., out of phase) makes them add constructively (see Supplementary Note~\ref{note:dual_rail_noise_analysis}). This also means that microwave crosstalk between a drive line and the non-targeted KO reduces the net interaction strength, since it drives the two modes in phase. The multi-path coupling design of Fig.~\ref{fig1}\textbf{a} balances the direct and indirect coupling paths to suppress this crosstalk (Supplementary Note~\ref{note:device}). Experimentally, we measure $-20$~dB of crosstalk between the drive lines and the KOs from the AC Stark shift, using the same two-tone spectroscopy as in Fig.~\ref{fig1}\textbf{d}.

\begin{figure*}[!htbp]
    \includegraphics[]{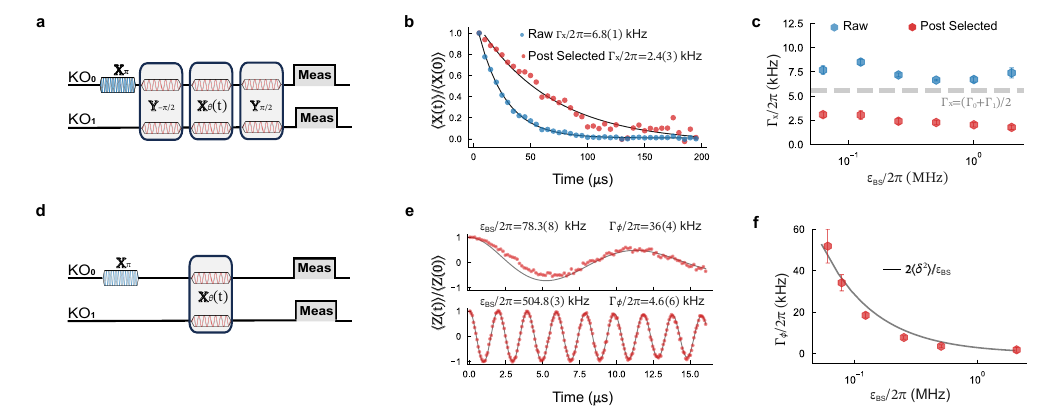}
    \caption{\label{fig4}\textbf{Dynamical protection of the dual-rail qubit}. 
    \textbf{a,} The pulse sequence for characterizing  spin-locking in the dual-rail qubit as discussed in the main text. \textbf{b,} The decay of the observable $\langle X(t)\rangle$ normalized to the initial value. The improvement with post-selection demonstrates the erasure bias due to the dynamical protection. \textbf{c,} The decay rate of the dual-rail qubit along the spin-locking axis as a function of the drive amplitude. The raw data decay rate is comparable to the average decay rate of the KOs. The post-selected decay rate saturates around 2 kHz which is $\sim4\times$ smaller. \textbf{d,} The pulse sequence for characterizing the dynamical protection of qubit coherence against low-frequency noise. \textbf{e,} Two examples of the post-selected Rabi oscillations at different beam-splitter drive amplitudes. At weak driving, spin-locking is nearly broken, and the evolution is barely sinusoidal. However, as the drive amplitude increases, the system becomes more robust, and the contrast of the Rabi oscillations increases. The fit includes the exponential decay obtained from panel \textbf{c} and the power-law decay coefficient $\Gamma_{\phi}$ using Eq.~\ref{eq:lf_noise}. \textbf{f,} The drop in the effect of low-frequency noise with the drive amplitude is consistent with the suppression of quasistatic noise due to spin-locking. From the fit we estimate $\sqrt{\langle \delta^2\rangle}/(2\pi)=38(1) \, \text{kHz}$ }
\end{figure*}
\vspace{0.5em}

Fig.~\ref{fig3}\textbf{d} and \ref{fig3}\textbf{e} show the calibration of Rabi oscillations in the dual-rail qubit driven via the beam-splitter interaction. To enhance calibration accuracy, we post-select on the absence of single-photon loss events at the end of each run. In Fig.~\ref{fig3}\textbf{d}, we present these Rabi oscillations as a function of time and drive detuning relative to the KOs frequency difference. The absence of a significant Stark shift from the nominal resonance ($\omega_{\text{d}}=\Delta_{\text{KO}}$) stems from the Stark shift being equal on both KOs with simultaneous driving.

The relative phase calibration while simultaneously driving both KOs is shown in Fig.~\ref{fig3}\textbf{e}. We observe the fastest Rabi oscillations when the displacements are out of phase. In this case, the contributions to $\epsilon_{\text{BS}}$ add up (top panel). On the other hand, driving with the same displacement  nearly cancels the beam-splitter interaction (bottom panel).  For clarity, the constant phase offset resulting from unequal electrical line delays has been subtracted.  The high contrast of these Rabi oscillations underscores the dynamical protection conferred by spin-locking to the charge drive. In the next section, we examine the nature of this dynamical protection and the performance of the dual-rail qubit using Clifford randomized benchmarking.

\begin{figure*}[t]
    \includegraphics[]{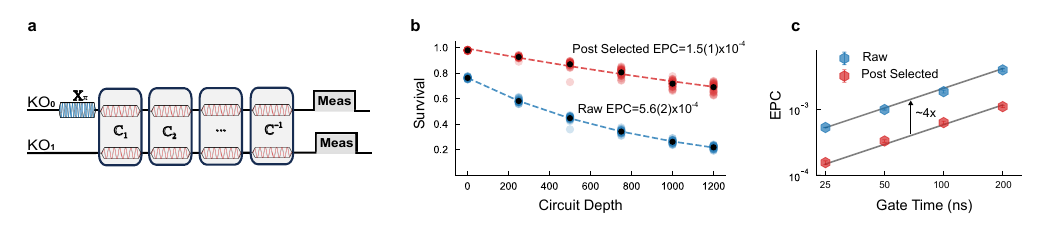}
    \caption{\label{fig5}\textbf{Randomized benchmarking of the dual-rail qubit.} 
    \textbf{a,} The protocol to estimate the EPC using randomized benchmarking of the dual-rail qubit relies on the survival of the quantum state after applying $N$ random Clifford gates (circuit depth) followed by the inverse gate. For each depth, we perform 30 random circuits. \textbf{b,} The survival of the quantum state for each circuit run is shown in the blue for the raw data and red for the post-selected data. The EPC is estimated from the exponential fit of the survival probability in each case. \textbf{c,} The EPC versus the gate time shows the same $4\times$ erasure bias maintained with faster gates.}
\end{figure*}
\vspace{0.5em}

\noindent\textbf{Dynamical protection with charge driving}

\noindent The dynamical protection of quantum systems is a powerful approach to preserve quantum information from noise. In particular, the saturation of two-level system (TLS) defects in the presence of strong fields and the spin-locking to a drive axis that does not commute with the noise axis are two common examples of this phenomenon \cite{Flux_noise_spectroscopy_NatPhys,Motional_Narrowing_ScienceAdvances,Decoherence_Superconducting_Qubit_PhysRevB}. The latter approach has been studied extensively in superconducting circuits with a variety of qubit spectroscopy techniques to reveal the underlying noise spectrum \cite{Hime2006_Science,Flux_noise_spectroscopy_NatPhys,Driven_dynamics_and_rotary_echo_PRL,Rotating_frame_relaxation_NatComm}. However, in practice, this protection can be accompanied by additional parasitic effects like enhanced decay and heating limiting its utility \cite{nguyen2024floquet,Fluxonium_Floquet_PhysRevApplied}.

In our system, this dynamical protection manifests in two ways: first, in the presence of strong Rabi driving, the action of single-photon loss inside the code space is suppressed due to the strong mixing of the KOs (i.e., $4\epsilon_{\text{BS}}>|\Gamma_0-\Gamma_1|$ as explained in Supplementary Note~\ref{note:dual_rail_noise_analysis}). Second, spin-locking mitigates the impact of low-frequency noise. This spin-locking protection arises because the drive and noise couple to non-commuting operators under the condition $\epsilon_{\text{BS}}^2 \gg \langle \delta^2 \rangle$, where $\delta$ denotes the effective dual-rail qubit detuning induced by the noise. More explicitly, projecting the beam-splitter Hamiltonian Eq.~\ref{eq:BS_Hamiltonian} together with the KOs detunings into the dual-rail basis yields the effective Hamiltonian:

\begin{equation}
    \hat{\mathcal{H}}_\text{eff}= \hat{P}_{DR}(\hat{\mathcal{H}}_\text{BS}+\Sigma_i \delta_i \hat{a}_i^{\dagger}\hat{a}_i)\hat{P}_{DR}=\epsilon_{\text{BS}}\hat{X}+\delta \hat{Z}
    \label{eq:eff_Hamiltonian}
\end{equation}

Where $\hat{P}_{DR}$ is the projection operator onto code space $\{ \ket{10} ,\ket{01} \}$, and the effective dual-rail detuning $\delta=(\delta_0-\delta_1)/2$. In this pseudo-spin picture, the energy splitting is $2\epsilon_{\text{BS}}$ in the absence of detuning. In practice, fluctuation in the detuning at this frequency (i.e., $\delta(\omega=2\epsilon_{\text{BS}})$) causes random bit-flips, and low-frequency noise in the detuning (i.e., $\delta(\omega\ll 2\epsilon_{\text{BS}})$) causes dephasing \cite{Decoherence_Superconducting_Qubit_PhysRevB}. Both effects are expected to decrease as $\epsilon_{\text{BS}}$ increases because the noise typically follows a $1/f$ spectrum.

Fig.~\ref{fig4}\textbf{a} shows the pulse sequence used to study spin-locking in dual-rail encoding. First, we excite one of the KOs to initialize the qubit along the $z$-axis, then we apply a logical $Y(-\pi/2)$ to prepare the qubit along the spin-locking axis. Next, we apply the spin-locking drive for a variable duration, and finally we apply a logical $Y(\pi/2)$ to rotate the qubit back to the $z$-axis and use simultaneous dispersive readout of the KOs to probe the outcome of each run. A typical spin-locking decay of the logical observable $\langle X(t)\rangle$ is shown in Fig.~\ref{fig4}\textbf{b}. The raw data show a typical exponential decay toward the ground state $\ket{00}$. However, when we keep only the experimental runs in which the system remains in the code space (i.e., post-selecting), we observe a much slower decay of the logical observable. This means our dual-rail qubit is erasure-biased along the spin-locking axis. This erasure bias remains fairly robust ($\sim4\times$) over a wide range of drive strengths, as shown in Fig.~\ref{fig4}\textbf{c}. The limit on the raw data is set by the average lifetime of the KOs. The decay of the  post-selected data saturates around 2 kHz. This limit  is similar to other experiments using the dual-rail encoding when only one erasure check is used \cite{levine2024dualrail}, since a single check cannot capture a decay followed by a heating event. With a dedicated ancilla mode for the mid-circuit erasure check, we expect this decay rate to be substantially lower.

To investigate protection against low-frequency noise, we examine the decay rate of the dual-rail qubit along the $z$-axis—orthogonal to the spin-locking drive axis. Experimentally, this rate is extracted from the decay of logical Rabi oscillations using the pulse sequence shown in Fig.~\ref{fig4}\textbf{d}. The origin of this protection becomes clear by expanding the energy splitting associated with Eq.~\ref{eq:eff_Hamiltonian}: $\Delta E=2\sqrt{ \epsilon^2_{\text{BS}}+\delta^2}\approx 2\epsilon_{\text{BS}}+\delta^2/\epsilon_{\text{BS}}$. In the strong-driving regime ($\epsilon_{\text{BS}} \gg \delta$), this expression highlights two key features: the qubit is sensitive to noise only to second order in $\delta$, and the overall impact of fluctuations is suppressed according to $1/\epsilon_{\text{BS}}$.

Fig.~\ref{fig4}\textbf{e} displays the post-selected Rabi oscillations for two distinct drive strengths. At low drive strength, spin-locking is weakly established, resulting in heavily damped, barely sinusoidal decay. However, as the drive strength increases, the system exhibits enhanced robustness, leading to high-contrast and prolonged Rabi oscillations. In both cases, the fit envelope $W(t)=e^{-\Gamma_x t} W_{\text{lf}}(t)$ is modeled as the product of the exponential decay extracted from panel \textbf{c} and the power-law decay characteristic of quasistatic second-order noise (see Supplementary Note~\ref{note:dual_rail_noise_analysis} for further details). The low-frequency contribution can be expressed as \cite{Decoherence_Superconducting_Qubit_PhysRevB,kubica2023erasure}:

\begin{equation}
    W_{\text{lf}}(t)\approx \frac{1}{[1+(\Gamma_\phi t)^2]^{1/4}}
    \label{eq:lf_noise}
\end{equation}

With the rate $\Gamma_\phi=2\left \langle \delta^2 \right \rangle /\epsilon_{\text{BS}}$. This inverse relationship is consistent with the data shown in Fig.~\ref{fig4}\textbf{f}. Around the value $\epsilon_{\text{BS}}/2\pi\approx 2 \, \text{MHz}$ the effect of low-frequency noise becomes negligible compared to the exponential decay rate of 2 kHz shown in panel \textbf{c}.

Leveraging this dynamical protection, we perform Clifford randomized benchmarking (CRB) to estimate the error per Clifford (EPC) gate  and the erasure bias of the dual-rail qubit, demonstrating performance comparable to the state of the art in static architectures despite using only one erasure check \cite{2_designs_PhysRevA,hashim2025benchmarking,levine2024dualrail}. The CRB protocol is shown in Fig.~\ref{fig5}\textbf{a}. We start by preparing the qubit along the $z$-axis by exciting one of the KOs. Next, we sample $N$ random Clifford gates followed by the inverse gate and measure both KOs. In Fig.~\ref{fig5}\textbf{b}, we show the exponential decay with the gate depth when performing the logical $X(\pi/2)$ gate in 25~ns achieving an EPC of $1.5\times 10^{-4}$ with post-selection. The EPC is estimated as $\frac{1}{2}(1-f)$, from the exponential fit $A f^N+B$ of the survival \cite{2_designs_PhysRevA,hashim2025benchmarking}. The erasure bias of $\sim 4$ is consistent with the lifetime improvement due to the dynamical protection. In Fig.~\ref{fig5}\textbf{c}, we show the EPC versus the gate time, which continues to improve with faster gates while maintaining the same erasure bias. We attribute this continuous improvement to the efficiency of the wave-mixing in our system of far-detuned strongly coupled KOs.

\vspace{0.5em}
\noindent\textbf{Discussion}

\noindent Our system demonstrates the utility of low-frequency charge driving for enabling diverse wave-mixing processes while providing dynamical protection against low-frequency noise. Using the AC Stark shift as a probe of the KO frequency sensitivity to charge driving, we showed that this sensitivity grows quadratically with both the drive amplitude and frequency. This scaling informed our design choices, favoring strongly coupled, far-detuned modes for efficient multi-mode operations. Within this architecture, we reset the KO in 82~ns to a residual population bounded by $0.7\%$, below its $2.5\%$ steady-state thermal population.

Using two KOs, we then encoded a dual-rail qubit and demonstrated both dynamical protection against low-frequency noise and a fourfold erasure bias with respect to single-photon loss. Leveraging this protection and the efficient wave mixing of our design, we implemented 25-ns logical gates with an error per Clifford of $5.6\times 10^{-4}$ ($1.5\times 10^{-4}$ after erasure post-selection) while preserving the erasure bias.

Two routes remain for improving these numbers. The raw error rate is set by the KO lifetime rather than by the drive scheme, whereas the post-selected rate saturates because a single erasure check at the end of the sequence cannot resolve a decay event followed by heating. Midcircuit erasure detection, implemented with a third KO as an ancilla, addresses the latter; such repeated checks were shown to be necessary for reaching a strong erasure bias in static-coupling architectures \cite{levine2024dualrail}. Because the protection reported here operates in a planar architecture with sub-GHz control tones, we also expect it to extend to beam-splitter-based entangling gates between dual-rail qubits \cite{Mehta2026Nature}.

While this proof-of-principle work confirms the viability of dynamical protection with sub-GHz charge tones, we expect it to be broadly applicable to more anharmonic oscillators. Ultimately, these results open new avenues for investigating drive protection mechanisms and mark a key step toward robust, high-fidelity superconducting quantum processors.

\section*{Acknowledgments}
We thank Chuan-Hong Liu, Noah Goss, and Ravi Naik for valuable discussions on the experiment. This work was supported by the U. S. Army Research Office under grant W911NF-22-1-0258.

\bibliographystyle{naturemag}
\bibliography{manuscript_samp}

\pagebreak
\widetext{}
\begin{center}
\textbf{\large Supplemental Materials}
\end{center}
\vspace{0.5em}

\newcounter{suppnote}
\renewcommand{\thesuppnote}{S\arabic{suppnote}}
\newcommand{\suppnote}[2]{%
  \refstepcounter{suppnote}%
  \noindent\textbf{Supplementary Note \thesuppnote\ – #1}%
  \label{#2}%
}
\setcounter{equation}{0}
\setcounter{figure}{0}
\setcounter{table}{0}
\renewcommand{\theequation}{S\arabic{equation}}
\renewcommand{\thefigure}{S\arabic{figure}}
\renewcommand{\thetable}{S\arabic{table}}
\renewcommand{\theHequation}{S\arabic{equation}}
\renewcommand{\theHfigure}{S\arabic{figure}}
\renewcommand{\theHtable}{S\arabic{table}}
\suppcitesbegin

\suppnote{2D chip design and circuit modes}{note:device}

\begin{figure}[tp]
    \includegraphics[width=0.5\textwidth]{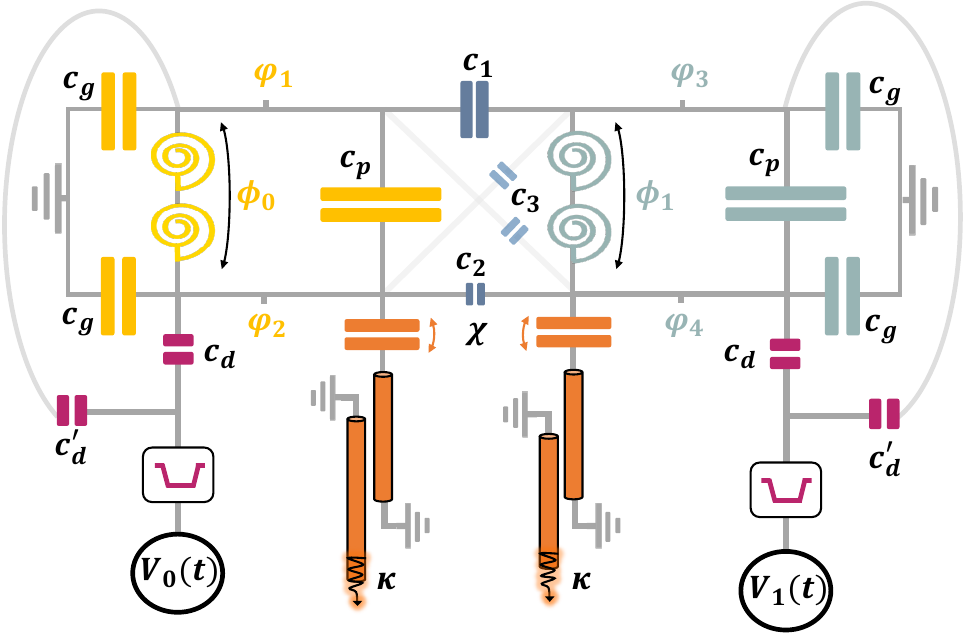}
    \caption{\label{fig1_supp}\textbf{2D chip circuit diagram}. Circuit diagram for the 2D chip showing every node in the floating design and the KO differential modes $\phi_0$, $\phi_1$. The coupling capacitors $c_1,c_2,c_3$ are shown in blue, the readout resonators and Purcell filters are shown in orange, and the charge pump lines are shown in magenta.}
\end{figure}

\noindent In this note, we provide more details on the device design parameters and trade-offs that achieve the performance presented in the main text, and we allocate the next notes to explore the main components in more detail. To start, we refer to the circuit diagram in Fig.~\ref{fig1_supp} and the corresponding node variables $\varphi_i$ to understand the chip design presented in Fig.~\ref{fig1}. Two KOs are realized as two differential modes $\phi_i$ of the circuit (teal and yellow). The nonlinear inductance of the KOs is realized using a double-SNAIL arrangement of Josephson junctions as shown in the diagram. The readout resonators and Purcell filters are shown as transmission lines with dispersive shift $\chi$ and decay rate $\kappa$. The control lines are capacitively coupled to the KOs with the applied voltage $V_i$.

For simplicity, we model the coupling between the KOs using three capacitors ($c_1$, $c_2$, and $c_3$) chosen to achieve a strong coupling rate of $g_{0,1}/2\pi \approx 90\text{ MHz}$ alongside a microwave crosstalk suppression of around $-20\text{ dB}$. We also assume that the two SNAILs in each KO are identical, allowing us to neglect the intermediate node. This assumption also requires the SNAIL capacitance to be small relative to the KOs' shunt capacitance. To analyze the eigenmodes of the circuit and the microwave crosstalk, we start with the full Lagrangian written in terms of the node flux variables $\varphi_i$ (setting $\Phi_0 = 1$):

\begin{equation}
    \mathcal{L} = \mathcal{L}_0 + \mathcal{L}_1 + \mathcal{L}_{\text{int}}
\end{equation}

\begin{equation}
    \mathcal{L}_0 = \frac{1}{2} c_{p} (\dot{\varphi}_1 - \dot{\varphi}_2)^2 + \frac{1}{2} c_{g} (\dot{\varphi}_1^2 + \dot{\varphi}_2^2) + \frac{1}{2} c_{d} (V_0 - \dot{\varphi}_2)^2+ \frac{1}{2} c^{'}_{d} (V_0 - \dot{\varphi}_1)^2 - U_{\text{SNAIL}}(\varphi_1 - \varphi_2)
\end{equation}

\begin{equation}
    \mathcal{L}_1 = \frac{1}{2} c_{p} (\dot{\varphi}_3 - \dot{\varphi}_4)^2 + \frac{1}{2} c_{g} (\dot{\varphi}_3^2 + \dot{\varphi}_4^2) + \frac{1}{2} c_{d} (V_1 - \dot{\varphi}_4)^2+ \frac{1}{2} c^{'}_{d} (V_1 - \dot{\varphi}_3)^2 - U_{\text{SNAIL}}(\varphi_3 - \varphi_4)
\end{equation}

\begin{equation}
    \mathcal{L}_{\text{int}} = \frac{1}{2} c_{1} (\dot{\varphi}_1 - \dot{\varphi}_3)^2+ \frac{1}{2} c_{2} (\dot{\varphi}_2 - \dot{\varphi}_4)^2 + \frac{1}{2} c_{3} (\dot{\varphi}_1 - \dot{\varphi}_4)^2  + \frac{1}{2} c_{3} (\dot{\varphi}_2 - \dot{\varphi}_3)^2
\end{equation}

From simulation we find that $c_3 \approx c_2$ in the final design but we keep it as a free parameter for now. Now defining the differential $\phi_n$ and common-mode $\theta_n$ phase variables as:

\begin{equation}
    \begin{aligned}
        \phi_0 = \varphi_1 - \varphi_2 \\
        \phi_1 = \varphi_3 - \varphi_4 \\
        \theta_0 = \varphi_1 + \varphi_2 \\
        \theta_1 = \varphi_3 + \varphi_4
    \end{aligned}
\end{equation}

Writing the Lagrangian in terms of the differential and common modes, we obtain:

\begin{equation}
    \mathcal{L}_{\phi_0} = \frac{1}{2} \left( c_p +\frac{c_{g}}{2}  + \frac{ c_{d\theta}}{2} + \frac{c_{1} +c_{2}+2c_3}{4} \right) \dot{\phi}_0^2 + \delta c_{d} V_0 \dot{\phi}_0 - U_{\text{SNAIL}}(\phi_0)
\end{equation}

\begin{equation}
    \mathcal{L}_{\phi_1} = \frac{1}{2} \left( c_p +\frac{c_{g}}{2}  + \frac{ c_{d\theta}}{2} + \frac{c_{1} +c_{2}+2c_3}{4} \right) \dot{\phi}_1^2 + \delta c_{d} V_1 \dot{\phi}_1 - U_{\text{SNAIL}}(\phi_1)
\end{equation}

\begin{equation}
    \mathcal{L}_{\theta_0} = \frac{1}{2}  \left( \frac{2 c_{g} + 2c_{d\theta} + c_{1} +  c_{2}+2c_3}{4} \right) \dot{\theta}_0^2 -  c_{d\theta} V_0 \dot{\theta}_0
\end{equation}

\begin{equation}
    \mathcal{L}_{\theta_1} = \frac{1}{2}  \left( \frac{2 c_{g} + 2c_{d\theta} + c_{1} + c_{2}+2c_3}{4} \right) \dot{\theta}_1^2 - c_{d\theta} V_1 \dot{\theta}_1
\end{equation}

with the effective capacitances:

\begin{equation}
    c_{d\theta} = \frac{c_{d} + c^{'}_{d}}{2}
\end{equation}

\begin{equation}
    \delta c_{d} = \frac{c_{d} - c^{'}_{d}}{2}
\end{equation}

The interaction Lagrangian can connect variables on the same side (local) or opposite sides (non-local) of the circuit, as shown in Fig.~\ref{fig1_supp}. For the local interaction, we get:

\begin{equation}
    \mathcal{L}_{\text{int},\text{local}} = -\frac{1}{2} \delta c_{d} \dot{\theta}_0 \dot{\phi}_0 - \frac{1}{2} \delta c_{d} \dot{\theta}_1 \dot{\phi}_1 + \frac{1}{4} c_{1} (\dot{\theta}_0 \dot{\phi}_0 + \dot{\theta}_1 \dot{\phi}_1) - \frac{1}{4} c_{2} (\dot{\theta}_0 \dot{\phi}_0 + \dot{\theta}_1 \dot{\phi}_1)
\end{equation}

The non-local interactions are mediated by $c_1$ and $c_2$, and we get:

\begin{equation}
    \mathcal{L}_{c_1, \text{int}} = -\frac{c_1}{4} ( \dot{\theta}_0+\dot{\phi}_0 )(\dot{\theta}_1 + \dot{\phi}_1) = -\frac{c_1}{4} ( \dot{\theta}_0 \dot{\theta}_1+\dot{\phi}_0 \dot{\phi}_1 + \dot{\theta}_0 \dot{\phi}_1 + \dot{\phi}_0 \dot{\theta}_1 )
\end{equation}

\begin{equation}
    \mathcal{L}_{c_2, \text{int}} = -\frac{c_2}{4} (\dot{\theta}_0-\dot{\phi}_0 )(\dot{\theta}_1 - \dot{\phi}_1) = -\frac{c_2}{4} (\dot{\theta}_0 \dot{\theta}_1+\dot{\phi}_0 \dot{\phi}_1   - \dot{\phi}_0 \dot{\theta}_1 - \dot{\theta}_0 \dot{\phi}_1)
\end{equation}

\begin{equation}
    \mathcal{L}_{c_3, \text{int}} = -\frac{c_3}{4} (\dot{\theta}_0-\dot{\phi}_0 )(\dot{\theta}_1 + \dot{\phi}_1)-  \frac{c_3}{4} (\dot{\theta}_0+\dot{\phi}_0 )(\dot{\theta}_1 - \dot{\phi}_1) = -\frac{c_3}{2} (\dot{\theta}_0 \dot{\theta}_1-\dot{\phi}_0 \dot{\phi}_1)
\end{equation}

We can see that by balancing the coupling (i.e., $c_1=c_2$ or relying on $c_3$) we can achieve the desired coupling between the differential modes without mixing the common and differential modes. However, in practice, it is hard to achieve the desired net coupling this way so we have to break this symmetry without sacrificing the crosstalk significantly. The common mode ($\theta_0$) can be eliminated to simplify the Lagrangian in terms of the remaining modes by enforcing the constraint $\frac{\partial \mathcal{L}}{\partial \dot{\theta}_0}=0 $:

\begin{equation}
    0=  c_{\theta} \dot{\theta}_0 - c_{d\theta} V_0 -\frac{1}{4} (2\delta c_{d}-c_1+c_2) \dot{\phi}_0 +\frac{1}{4} (c_2-c_1) \dot{\phi}_1 -\frac{1}{4} (c_1+c_2+2c_3) \dot{\theta}_1
    \end{equation}

\begin{equation}
    c_{\theta}=\frac{2 c_{g} + 2c_{d\theta} + c_{1} +c_{2}+2c_3}{4} \approx c_g/2
\end{equation}

So we can write the common mode in terms of the differential modes and the external drive as:
\begin{equation}
    \dot{\theta}_0 = \frac{1}{c_{\theta}} \left(c_{d\theta} V_0 + \frac{2\delta c_{d}-c_1+c_2}{4} \dot{\phi}_0 - \frac{c_2-c_1}{4} \dot{\phi}_1 + \frac{c_1+c_2+2c_3}{4} \dot{\theta}_1 \right)
\end{equation}

This introduces several small corrections to the Lagrangian, but we are primarily interested in the crosstalk it mediates between the applied voltages $V_n$ and the opposite mode of the circuit $\phi_n$. The effective Lagrangian then takes the form:

\begin{equation}
    \mathcal{L}_{\text{eff}}=\mathcal{L}_{\text{static}}+\mathcal{L}_{\text{d}}
\end{equation}

\begin{equation}
    \mathcal{L}_{\text{static}}=\frac{1}{2}c_{\text{KO}}\left({\dot{\phi}}_0^2+{\dot{\phi}}_1^2\right)-U_{\text{SNAIL}}\left(\phi_0\right)-U_{\text{SNAIL}}\left(\phi_1\right)-c_c{\dot{\phi}}_0{\dot{\phi}}_1
\end{equation}

\begin{equation}
    \mathcal{L}_{d} \approx \left(\delta c_d V_0- c_{d\theta}\frac{\delta c}{c_g}V_1\right){\dot{\phi}}_0+\left(\delta c_d V_1- c_{d\theta}\frac{\delta c}{c_g}V_0\right){\dot{\phi}}_1
\end{equation}

with the effective capacitances:
\begin{equation}
c_{\text{KO}} \approx c_p+\frac{c_{g}}{2}  + \frac{c_{d\theta}}{2} + \frac{c_{1} + c_{2}+2c_3}{4}
\end{equation}

\begin{equation}
c_c\approx\frac{c_1+c_2-2c_3}{4}
\end{equation}

\begin{equation}
\delta c\approx\frac{c_1-c_2}{2}
\end{equation}

For simplicity, we assume a single drive in the following treatment since the analysis is identical for the other drive ($V_0=V_d$ and $V_1=0$). To construct the Hamiltonian, we first compute the conjugate charges:

\begin{equation}
\mathrm{Q}_0=\frac{\partial \mathcal{L}}{\partial \dot{\phi}_0}=c_{\text{KO}} \dot{\phi}_0-c_c \dot{\phi}_1 +\delta c_d V_d \rightarrow \mathrm{Q}_0^{\prime}=c_{\text{KO}} \dot{\phi}_0-c_c \dot{\phi}_1
\end{equation}

\begin{equation}
    \mathrm{Q}_1=\frac{\partial \mathcal{L}}{\partial \dot{\phi}_1}=c_{\text{KO}} \dot{\phi}_1-c_c \dot{\phi}_0- c_{d\theta}\frac{\delta c}{c_g} V_d \rightarrow \mathrm{Q}_1^{\prime}=c_{\text{KO}} \dot{\phi}_1-c_c \dot{\phi}_0
    \end{equation}

\begin{equation}
    \vec{Q}' = \boldsymbol{C} \, \dot{\vec{\phi}}
    \quad \Rightarrow \quad
    \dot{\vec{\phi}} = \boldsymbol{C}^{-1} \vec{Q}'
\end{equation}

The Hamiltonian then reads:

\begin{equation}
    \mathcal{H} = \vec{Q}^T \vec{\dot{\phi}} - \mathcal{L} = \vec{Q}^T \boldsymbol{C}^{-1} \vec{Q}' - \frac{1}{2} \vec{Q}'^T \boldsymbol{C}^{-1} \vec{Q}' + U_{\text{SNAIL}}(\phi_0,\phi_1) -  V_d \begin{pmatrix} \delta c_d & -c_{d\theta}\frac{\delta c}{c_g} \end{pmatrix}  \boldsymbol{C}^{-1} \vec{Q}'
\end{equation}

The first two terms simplify to:

\begin{equation}
\vec{Q}^T \mathbf{C}^{-1} \vec{Q}' -\frac{1}{2} \vec{Q}'^T \mathbf{C}^{-1} \vec{Q}'=\frac{1}{2c_{\text{KO}}}\left(Q_0^2+Q_1^2\right) +\frac{c_c}{c_{\text{KO}}^2}\left(Q_0 Q_1\right)
\end{equation}

The last term simplifies to:
\begin{equation}
    \begin{aligned}
        -V_d \begin{pmatrix} \delta c_d & -c_{d\theta}\frac{\delta c}{c_g} \end{pmatrix} \mathbf{C}^{-1} \vec{Q}' &\approx  - \frac{V_d}{c^2_{\text{KO}}} \begin{pmatrix} \delta c_d & -c_{d\theta}\frac{\delta c}{c_g} \end{pmatrix}
    \begin{pmatrix} c_{\text{KO}} & c_c \\ c_c & c_{\text{KO}} \end{pmatrix} \begin{pmatrix} Q_0 \\ Q_1 \end{pmatrix} \\
    &= -V_d \frac{\delta c_d }{c^2_{\text{KO}}}(c_{\text{KO}} Q_0 + c_c Q_1) +  V_d \frac{c_{d\theta}}{c^2_{\text{KO}}} \frac{\delta c}{c_g} (c_c Q_0 + c_{\text{KO}} Q_1)
    \end{aligned}
\end{equation}

We thus identify the crosstalk Hamiltonian as:

\begin{equation}
    \mathcal{H}_\text{crosstalk}=-\frac{1}{c_{\text{KO}}} \left(\delta c_d \frac{c_c}{c_{\text{KO}}}-c_{d\theta}\frac{\delta c}{c_g}\right) V_d Q_1
    \label{eq:crosstalk_Hamiltonian}
\end{equation}

By tuning the coupling asymmetry, we can reduce the crosstalk between the bare oscillators. However, it is worth noting that even in the absence of bare-state crosstalk, there will be some crosstalk to the dressed states due to the net coupling between the oscillators. This crosstalk is proportional to the degree of hybridization between the modes ($\sim g_{0,1}/\Delta_{\text{KO}}$). Examining the leading drive term, we find:

\begin{equation}
    \hat{\mathcal{H}}_\text{d}=-\frac{\delta c_d V_d}{c_{\text{KO}}}\hat{Q}_0 = -i2e\frac{\delta c_d V_d}{c_{\text{KO}}}n_{\text{ZPF}} (\hat{a}_0-\hat{a}^{\dagger}_0) \rightarrow i\epsilon_d  (\hat{a}_0-\hat{a}^{\dagger}_0)+i\frac{g_{0,1}}{\Delta_{\text{KO}}}\epsilon_d (\hat{a}_1-\hat{a}^{\dagger}_1)
\end{equation}

\begin{figure}[tp]
    \includegraphics[width=0.99\textwidth]{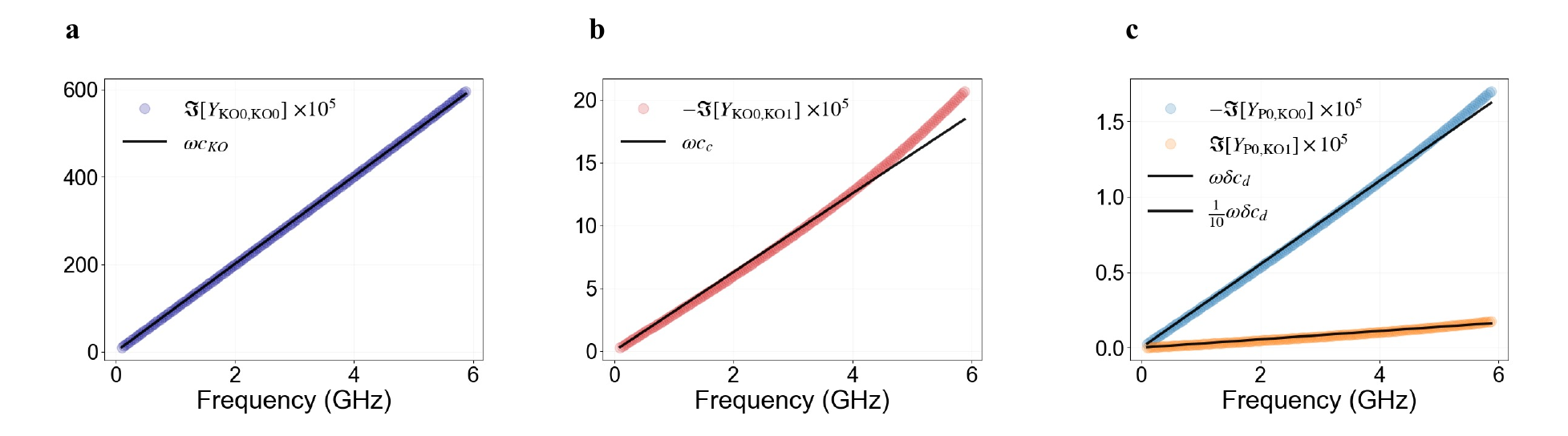}
    \caption{\textbf{KOs coupling and crosstalk simulation.} \textbf{a,} The diagonal part of the admittance matrix for the KOs gives the admittance of the net shunt capacitance, which is effectively $Y=j \omega c_{\text{KO}}$. From the fit, we estimate $c_{\text{KO}}= 160$ fF. \textbf{b,} The off-diagonal part of the admittance matrix between the KOs is effectively the admittance of the coupling capacitor $Y=-j \omega c_c$. From the fit, we estimate $c_c= 5$ fF. \textbf{c,} Similarly, the off-diagonal part between the pads of KO0 and its control line is proportional to $\delta c_d=0.44$ fF. This also allows us to simulate the crosstalk to KO1, which we find to be 1/10 of the coupling to the targeted KO with a net minus sign.}
    \label{fig:Coupled_KOs_simulation}
\end{figure}

We iterate over the design of the coupling capacitance and simulate the modes quality factor to approach the limit set by the ratio $g_{0,1}/\Delta_{\text{KO}}$. In Fig.~\ref{fig:Coupled_KOs_simulation}, we present some of the simulation results for this mask. First, we use driven modal simulation in ANSYS without the inductors to estimate the capacitive coupling between the KOs. Using lumped ports instead of inductors allows us to isolate the contribution from hybridization. In part \textbf{a}, the diagonal part of the admittance matrix for any of the KOs gives the admittance shunting the SNAILs, which is effectively $Y=j \omega c_{\text{KO}}$. From the linear trend, we estimate $c_{\text{KO}}=160$ fF. In part \textbf{b}, we see the off-diagonal part of the admittance matrix, which is effectively minus the admittance of the coupling capacitor $Y=-j \omega c_c$. From the fit, we estimate $ c_c = 5$ fF. This enables us to estimate the coupling between the KOs as $g_{0,1}\approx 90$~MHz.  In part \textbf{c}, we include a third port at the end of the charge control line for KO0 to simulate the capacitive coupling and the resulting crosstalk to KO1. From the fit, we estimate $\delta c_d=0.44$ fF and $10^{-1}$ crosstalk to the other KO. This crosstalk agrees well with the value predicted by Eq.~\ref{eq:crosstalk_Hamiltonian} for the design parameters of the chip in Table~\ref{tab:KOs_parameters}.

The experimental parameters for the 2 KOs chip are shown in Table \ref{tab:KOs_parameters}. The detuning between the KOs was less than 1 GHz due to fabrication mismatch and varied across cooldowns. However, since the KOs are tunable, we can maximize the detuning to study the AC Stark shift across a wider range. The tuning curve and nonlinearity of the SNAIL are treated in the next supplemental notes. The fabrication details are identical to the work in \cite{qing2026benchmarking} and follow the recipes developed and characterized in \cite{kreikebaum2020jjvariation,altoe2022localization}.

\newpage

\begin{table}[htb]

    \caption{Experimental parameters for the KOs circuit in Fig.~\ref{fig1_supp}}
    \centering
    \begin{tabular}{ |p{9cm}  | c | c |}
    \hline  \hline
     \textbf{Parameter} & \textbf{KO0} & \textbf{KO1}   \\ [1ex] 
    \hline    
    KO frequency in Dual-rail $\omega_{\text{KO}}/2\pi$& 6.333 GHz  &  5.703 GHz  \\ [1ex] 
    \hline 
    KO anharmonicity    $-\alpha/2\pi$ & -4.4 MHz &-7.0 MHz
         \\[1ex]
    \hline 
    KO charging energy $E_c/h$ & 0.121 GHz  &  0.121 GHz  \\ [1ex]
    \hline 
    Number of SNAILs& 2  & 2    \\ [1ex]
    \hline 
    KO single-photon decay time $T_1=\Gamma^{-1}$& 27.0 $\mu$s & 31.0 $\mu$s     \\ [1ex]
    \hline 
    KO Ramsey decay time $T_2^*$& 3.4 $\mu$s & 7.8 $\mu$s     \\ [1ex]
    \hline
    KO echo decay time $T_2^e$& 13.5 $\mu$s & 18.5 $\mu$s     \\ [1ex]
    \hline
    Readout resonator frequency $\omega_R/2\pi$& 7.56
    GHz & 6.92 GHz     \\[1ex]
    \hline 
    Readout resonator decay time  $\kappa^{-1}_R$& 1.0 $\mu$s & 1.1 $\mu$s     \\ [1ex]
    \hline 
    Purcell resonator frequency $\omega_P/2\pi$& 7.64
    GHz & 7.12 GHz     \\[1ex]
    \hline 
    Purcell resonator linewidth  $\kappa_P/2\pi$& 50 MHz & 40 MHz     \\ [1ex]
    \hline 
    KO--Resonator coupling  $g_{\text{KO,R}}/2\pi$ & \multicolumn{2}{c|}{110 MHz}     \\ [1ex]
    \hline 
    KO shunt capacitance $c_{\text{KO}}$ & \multicolumn{2}{c|}{160 fF}     \\ [1ex]
    \hline 
    KO pads mutual capacitance $c_p$ & \multicolumn{2}{c|}{79 fF}     \\ [1ex]
    \hline 
    KO capacitance to ground $c_g$ & \multicolumn{2}{c|}{146 fF}     \\ [1ex]
    \hline 
    KOs coupling capacitance $c_c$ & \multicolumn{2}{c|}{5 fF}     \\ [1ex]
    \hline 
    KOs coupling capacitance asymmetry $\delta c$ & \multicolumn{2}{c|}{10 fF}     \\ [1ex]
    \hline
    KO charge line coupling capacitance  $\delta c_d$& \multicolumn{2}{c|}{0.44 fF}     \\ [1ex]
    \hline 
    KO charge line total capacitance  $ c_{d\theta}$& \multicolumn{2}{c|}{0.8 fF}     \\ [1ex]
    \hline 
    \end{tabular}
    \label{tab:KOs_parameters}
    \end{table}

\newpage
\suppnote{Control line Purcell decay and on-chip filter}{note:control_line_Purcell_decay}

\begin{figure}[pt!]
    \includegraphics[width=0.99\textwidth]{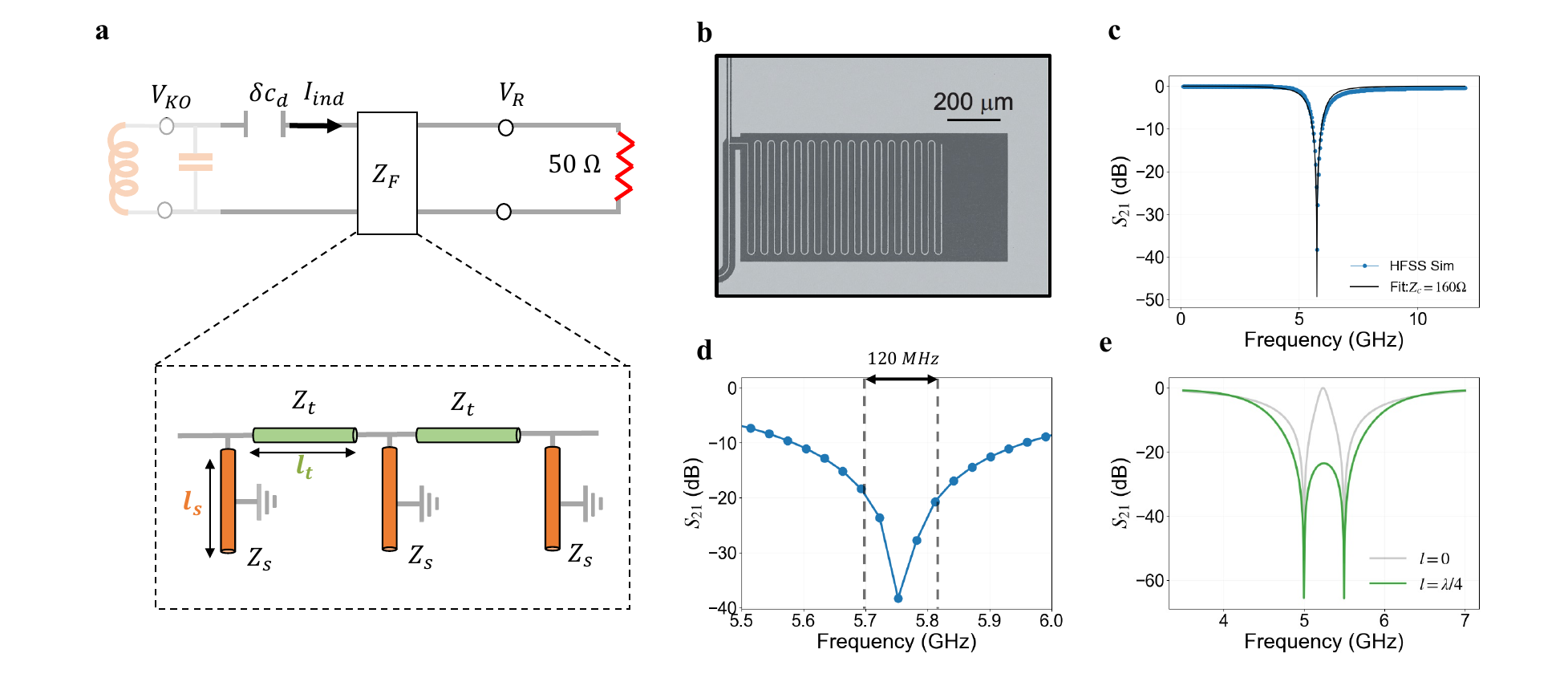}
    \caption{\textbf{Purcell decay and band-stop filter design.} \textbf{a,} The Purcell decay of the KO can be analyzed through a simple LC model where the oscillating mode voltage induces a current $I_{\text{ind}}$ proportional to the coupling capacitance $\delta c_d$ which is dissipated at the $50\, \Omega$ load resistor. To suppress this effect, we insert an impedance $Z_F$ that acts as a short to ground at the qubit frequency and as an open far off-resonance. This can be accomplished with an arrangement of stubs as shown in the subfigure. ~\textbf{b,} A single stub consists of a meander with an open boundary condition, with total length $l \approx \lambda/4$ at the qubit frequency. \textbf{c,} The simulated transmission coefficient of the single stub filter shows a stop band around 5 GHz. \textbf{d,} A zoomed-in view of the stop band shows a 20 dB suppression over a 120 MHz bandwidth. \textbf{e,} The transmission of a two-stub filter design with two values for the length of the series stub (i.e., impedance inverter). This design allows for a much wider stop-band bandwidth.}
    \label{fig:Filter_simulation}
\end{figure}

\noindent The capacitive coupling between the KOs and the control lines used to activate the wave mixing processes comes with an inherent Purcell decay. As in previous experiments \cite{hajr2024kerrcat2d,qing2026benchmarking}, we maintain strong capacitive coupling to increase the efficiency of far-off-resonance wave mixing processes and mitigate resonant Purcell decay using an on-chip band-stop filter. To start, we examine the Purcell decay through the simple model presented in Fig.~\ref{fig:Filter_simulation}\textbf{a}, where the oscillating voltage of the qubit $V_{\rm KO}$ induces a current that dissipates into the $R=50 \, \Omega$ load resistor modeling the transmission line. Because the impedance of the coupling capacitance is much larger than the load resistor at the qubit frequency, the induced current has the form $I_{\text{ind}}=j\omega_{\rm KO} c{'} V_{\rm KO}$ which dissipates a power $P_{\text{diss},0}=\frac{1}{2}R \left|I_{\text{ind}}\right|^2$ limiting the quality factor to:

\begin{equation}
    Q_e=\omega_{\rm KO}\,\frac{\tfrac{1}{2}c_{\rm KO}V_{\rm KO}^2}
    {\tfrac{1}{2}R\left(\omega_{\rm KO}\,\delta c_d\,V_{\rm KO}\right)^{2}}
    =\frac{c_{\rm KO}}{\delta c_d^{2}R\,\omega_{\rm KO}},
\end{equation}

For the parameters given in Table \ref{tab:KOs_parameters}, we estimate $Q_e\approx 4 \times 10^5$, which limits the lifetime of the KO to $11 \, \mu$s. To overcome this limitation, we use the band-stop filter shown in Fig.~\ref{fig:Filter_simulation}\textbf{a} with impedance $Z_{\text{F}}$ that acts as a short to ground at the qubit frequency and as an open far off-resonance. Around the qubit frequency, we have the following hierarchy of impedances: $\left|Z_{\delta c_d}\right|\gg R\gg |Z_F|$, so the voltage across the load resistor is:

\begin{equation}
    V_R\approx\frac{V_{\rm KO}}{|Z_{\delta c_d}|}|Z_F|
\end{equation}

As a result, the power dissipated in the 50 $\Omega$ resistor is:

\begin{equation}
    P_{\text{diss}}^\prime=\frac{|V_R|^2}{2R}\approx\frac{1}{2R}\ \left|\frac{V_{\rm KO}}{Z_c}Z_F\right|^2\ =\left(\frac{1}{2}R\left|I_{\text{ind}}\right|^2\right)\frac{\left|Z_F\right|^2}{R^2}=P_{\text{diss},0}\frac{\left|S_{21}\right|^2}{4}
\end{equation}

The dissipated power is therefore suppressed by $|S_{21}|^2$, the transmission coefficient of the filter measured in the 50 $\Omega$ transmission line, together with an additional factor of $1/4$ that arises because the transmission coefficient is referenced to the 50 $\Omega$ characteristic impedance rather than to the qubit impedance, which is much larger.

Our approach to realize this filter uses multiple transmission-line sections, as shown in the schematic in Fig.~\ref{fig:Filter_simulation}\textbf{a}. A parallel $\lambda/4$ stub at the qubit frequency acts as a short, and a $\lambda/4$ stub in series acts as an impedance transformer, allowing us to increase the bandwidth of the filter. More precisely, the input impedance of an open stub at distance $l$ is:

\begin{equation}
    Z(l)=\frac{V(l)}{I(l)}=Z_{s,c}\,\frac{e^{jkl}+\Gamma e^{-jkl}}{e^{jkl}-\Gamma e^{-jkl}}\;\xrightarrow[\ \Gamma=1\ ]{}\; Z_s=-jZ_{s,c}\cot\!\left(\frac{2\pi}{\lambda}l\right),
\end{equation}

The chip layout of this stub is shown in Fig.~\ref{fig:Filter_simulation}\textbf{b} and the simulated transmission coefficient is shown in Fig.~\ref{fig:Filter_simulation}\textbf{c} using \cite{pozar2011microwave}:

\begin{equation}
    S_{21}=\frac{1}{1+Z_0/(2Z_s)}=\frac{1}{1+\frac{jZ_0}{2Z_{s,c}}\tan\left(\frac{\pi}{2}f/f_s\right)}
    \label{eq:quarter_wave_stub_S21}
\end{equation}

with $f_s= \nu /4l$ being the quarter-wavelength frequency. The stub's relatively high impedance of $160 \, \Omega$ limits the 20 dB bandwidth to 120 MHz. To increase the bandwidth, we can add multiple stubs with impedance transformers to avoid destructive interference between them. To start, let us consider two stubs with center frequencies $f_1$ and $f_2$ in parallel, with zero spacing between them. The effective impedance close to the center frequencies is given by: 

\begin{equation}
\frac{1}{Z_{\text{eff}}}=\frac{1}{Z_{s1}}+\frac{1}{Z_{s2}}=\frac{2/\pi}{jZ_{s,c}}\left(\frac{f_1}{\delta f_1}+\frac{f_2}{\delta f_2}\right)
\end{equation}

With $\delta f_i = f-f_i$ being the detuning from the center frequency $f_i$. For $f\approx(f_1+f_2)/2$, the two stubs anti-resonate into an open circuit and the filter stops working between the two stop bands. To prevent this destructive interference, we can insert a $\lambda/4$ series transmission-line section that acts as an impedance inverter. Fig.~\ref{fig:Filter_simulation}\textbf{e} shows the simulation for a two-stub filter design with two values for the length of the impedance inverter. To derive the effect of the impedance inverter, we can look at the ABCD matrix of this system around the middle frequency:

\begin{equation}
    ABCD\approx\left(\begin{matrix}1&0\\\frac{1}{Z_{s2}}&1\\\end{matrix}\right)\ \left(\begin{matrix}0&jZ_{t,c}\\\frac{j}{Z_{t,c}}&0\\\end{matrix}\right)\left(\begin{matrix}1&0\\\frac{1}{Z_{s1}}&1\\\end{matrix}\right)=\left(\begin{matrix}\frac{jZ_{t,c}}{Z_{s1}}&jZ_{t,c}\\\frac{jZ_{t,c}}{Z_{s1}Z_{s2}}+\frac{j}{Z_{t,c}}&\frac{jZ_{t,c}}{Z_{s2}}\\\end{matrix}\right)
\end{equation}
From which we can evaluate the transmission coefficient as: 

\begin{equation}
    S_{21}=\frac{-2j}{\dfrac{Z_{t,c}}{Z_0}+\dfrac{Z_0}{Z_{t,c}}
    +Z_{t,c}\!\left(\dfrac{1}{Z_{s1}}+\dfrac{1}{Z_{s2}}\right)+\dfrac{Z_0Z_{t,c}}{Z_{s1}Z_{s2}}}
\end{equation}

In the denominator, the second and third terms reproduce what we would get without the series stub, which cancels near the middle frequency. The series stub introduces the final term, so the contributions from the filters therefore add in decibels. The performance of the full filter design is shown in reference \cite{hajr2024kerrcat2d,hajr_thesis}.

\newpage

\suppnote{SNAIL nonlinearity analysis}{note:snail_nonlinearity}

\begin{figure}[pt!]
    \includegraphics[width=0.99\textwidth]{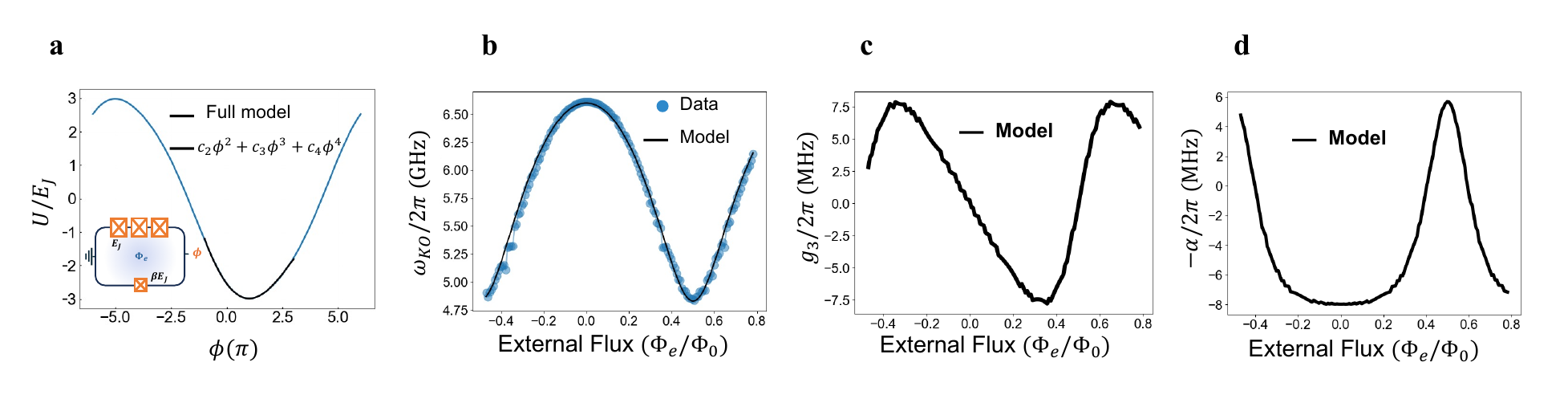}
    \caption{\textbf{SNAIL potential and nonlinearity.} \textbf{a,} The SNAIL consists of three large junctions on one side and a small junction on the other side, with a ratio $\beta \approx 0.1$ and an external flux bias around $\Phi_{e}=\Phi_0/3$. Near the potential minima, the Taylor expansion includes a non-zero cubic term due to the potential's asymmetry. \textbf{b,} The experimental tuning curve of a typical double-SNAIL KO with similar parameters to Table \ref{tab:KOs_parameters}. By matching the tuning curve, we can predict the KO nonlinearities. \textbf{c,} The third-order nonlinearity as a function of the external flux. This nonlinearity is maximized around $\Phi_{e}=\pm\Phi_0/3$ and vanishes at the sweet spots. \textbf{d,} The predicted anharmonicity of the KO. Due to destructive interference between the SNAIL branches, the anharmonicity is much smaller than the KO charging energy.}
    \label{fig:SNAIL}
\end{figure}

\noindent The use of the Josephson junctions in the SNAIL arrangement renders the charge dispersion negligible because the equivalent single junction oscillator (i.e., one that has the same frequency and anharmonicity at zero flux) has a ratio $E_J/E_c=\mathcal{O}(10^5)$. This enables us to focus on the wave mixing processes associated with the KO nonlinearity. Moreover, the third-order nonlinearity triggers the beam-splitter interaction with a single photon from the drive. Fig.~\ref{fig:SNAIL} shows the details of the SNAIL and the SNAIL-based KO used in the first generation of devices (uncoupled KOs chip). In part \textbf{a}, we see the circuit of the SNAIL consisting of three large junctions on one side and a small junction on the other side, with external flux threading the loop. The potential associated with the SNAIL has the form \cite{frattini2017snail,frattini2018snailpam}:

\begin{equation}
    U_{\text{SNAIL}}=-\beta E_J\cos{\left(\phi\right)}-3E_J\cos{\left(\frac{\phi_{e}-\phi}{3}\right)}=E_J\sum_{n=2}^{\infty} \frac{a_n}{n!}\tilde{\phi}^n
\end{equation}

where $\tilde{\phi}$ is the reduced phase variation away from the minima and $\beta$ is the ratio between the small and large junctions and $\phi_{e}$ is the reduced external flux threading the loop ($\phi_{e}=2\pi \Phi_{e}/\Phi_0$). Unlike the SQUID, the SNAIL potential does not have a simple analytic expansion; instead, it is typically expanded in a Taylor series around the minima with coefficients $a_n$. When the SNAIL is capacitively shunted, the resulting KO has the Hamiltonian:

\begin{equation}
{\hat{\mathcal{H}}}_{\text{SNAIL}}\approx \omega_{\text{KO}}{\hat{a}}^\dag\hat{a}+g_3\left({\hat{a}}^\dag+\hat{a}\right)^3+g_4\left({\hat{a}}^\dag+\hat{a}\right)^4 
\end{equation}

where we have used $\omega_{\text{KO}}=\sqrt{8E_cE_J a_2}$ and expressed the nonlinearity in terms of the bosonic operators. We have neglected higher-order terms since they are suppressed by $\phi_{\text{ZPF}}=(2E_c/E_J a_2)^{1/4}$. The coefficients of the third-order and fourth-order nonlinearities, in terms of the coefficients of the Taylor expansion of the potential, are given by:

\begin{equation}
    g_3=\frac{1}{6} \frac{a_3}{a_2} \sqrt{E_c  \omega_{\text{KO}}}
\end{equation}

\begin{equation}
    g_4=\frac{a_4}{12 a_2} E_c 
\end{equation}

The anharmonicity ($-\alpha$) is approximately twice the coefficient of $\hat{a}^{\dag2}\hat{a}^2$ in the Hamiltonian which leads to:

\begin{equation}
    \alpha=-12(g_4-5g_3^2/\omega_{\text{KO}})
\end{equation}

The second term is the leading correction due to the cubic term, which can be calculated as follows:

\begin{equation}
\begin{aligned}
    E_n^{(2)} &= \sum_{m \neq n} \frac{|\langle m|H_3|n \rangle|^2}{E_n - E_m} = g_3^2 \sum_{m \neq n} \frac{|\langle m|(\hat{a}^\dagger + \hat{a})^3|n \rangle|^2}{(n-m)\omega_{\text{KO}}} \\[1em]
    &= g_3^2 \left( \frac{|\langle n+3|\hat{a}^{\dagger 3}|n \rangle|^2}{-3\omega_{\text{KO}}} + \frac{|\langle n-3|\hat{a}^3|n \rangle|^2}{3\omega_{\text{KO}}} \right. \\
    &\quad \left. + \frac{|\langle n+1|\hat{a}^{\dagger 2}\hat{a} + \hat{a}^\dagger\hat{a}\hat{a}^\dagger + \hat{a}\hat{a}^{\dagger 2}|n \rangle|^2}{-\omega_{\text{KO}}} + \frac{|\langle n-1|\hat{a}^2\hat{a}^\dagger + \hat{a}\hat{a}^\dagger\hat{a} + \hat{a}^\dagger\hat{a}^2|n \rangle|^2}{\omega_{\text{KO}}} \right) \\[1em]
    &= - 30 \frac{g_3^2}{\omega_{\text{KO}}} (n^2 - n) + \mathcal{O}(n)
    \end{aligned}
\end{equation}
where we have used the fact that the transition frequency between the Fock states is approximately $\omega_{\text{KO}}$.

We can see that the ratio $g_3/\alpha$ is proportional to $\sqrt{\omega_{\text{KO}}/E_c}$, which makes it desirable to reduce $E_c$ as much as possible. However, to avoid using capacitor pads with very large footprints, we can use multiple SNAILs in series to dilute the nonlinearities. This approach, however, reduces $g_3$, the nonlinearity we seek to maximize, but it increases the ratio $g_3/\alpha$. In principle, with $M$ identical SNAILs and assuming we scale $E_J$ to keep the frequency constant, the nonlinearities scale as: 

\begin{equation}
    g_3 \rightarrow  g_3/M \quad \text{and} \quad \alpha \rightarrow  \alpha/M^2
\end{equation}

Also, practically, the presence of small parasitic inductance in series with the SNAIL increases the total inductance and dilutes the nonlinearity according to the scaling \cite{frattini2018snailpam}:

\begin{equation}
    \begin{aligned}
        &a_2 \rightarrow p \times a_2 \\
        &a_3 \rightarrow p^3 \times a_3 \\
        &a_4 \rightarrow p^4 \times (a_4 -\frac{3 a_3^2}{a_2} (1-p))
    \end{aligned}
\end{equation}

with $p=L_s/(L_s+L_p)$, the participation ratio of the SNAIL in the total inductance. In Fig.~\ref{fig:SNAIL} \textbf{a}, we show the potential of the SNAIL when biased at $\Phi_{e}=\Phi_0/3$ and the Taylor expansion of the potential up to the quartic term, which captures the potential around the minima. The potential is clearly asymmetric around the minima, a signature of the cubic term. In Fig.~\ref{fig:SNAIL} \textbf{b}, we report the experimental tuning curve of an early device with a double-SNAIL KO. The fit is constructed using eigenmodes of the effective Hamiltonian up to the fourth order for each flux point (i.e., for each flux point, we expand the SNAIL potential around the minima and use the coefficients to build a Hamiltonian that we diagonalize to obtain the lowest transition frequency). The same fit predicts a third-order nonlinearity, which we show in Fig.~\ref{fig:SNAIL} \textbf{c}. The third-order nonlinearity peaks around $\Phi_{e}=\pm\Phi_0/3$. In Fig.~\ref{fig:SNAIL} \textbf{d}, we see the predicted anharmonicity of the KO, which typically agrees up to $20 \%$ with the measured anharmonicity.

\newpage
\vspace{0.5em}

\suppnote{Wave mixing with low frequency charge driving}{note:wave_mixing}

\noindent In the absence of DC charge sensitivity in the KO, the remaining AC sensitivity is dictated by the nonlinearity of the oscillator. In this note, we derive the effect of a low-frequency charge driving on the KO using the induced displacement to leading order in the driving strength. Next, we examine the AC Stark shift induced by this drive and perform numerical simulation to verify our derivation. 

In this work, low-frequency charge driving refers to a drive that is far detuned from the KO resonance frequency relative to the anharmonicity  (i.e., $|\omega_{\text{KO}}-\omega_d| \gg \alpha$). In this case, the drive-induced displacement in the KO depends only on the drive frequency and detuning, not on the anharmonicity. First, let us consider a KO with weak nonlinearities subject to a generic single-photon drive (i.e., it could be phase or charge): 

\begin{equation}
    \hat{\mathcal{H}}_0=\omega_{\text{KO}}{\hat{a}}^\dag\hat{a}+\sum_{n\geq3}{g_n\left(\hat{a}+\text{h.c.}\right)^n}
    \end{equation}

    \begin{equation}
        \hat{\mathcal{H}}_d=2\cos{(\omega_d t)} (\epsilon_d \hat{a}^\dagger+\epsilon_d^*\hat{a})
        \label{eq:charge_drive}
    \end{equation}

To understand the effect of the drive, we can go to a rotating displaced frame as follows:

\begin{equation} 
    \hat{U}=\hat{R}_{\omega_r}\hat{D}_\xi=e^{i\omega_r t {\hat{a}}^\dag\hat{a}}\ e^{\xi{\hat{a}}^\dag-\xi^\ast\hat{a}}
\end{equation}

The displacement shifts the bosonic operators, and the rotation frame facilitates the RWA:
 
\begin{equation}
\hat{D}_\xi\hat{a}\hat{D}_\xi^\dag=\hat{a}-\xi
\end{equation}

\begin{equation}
\hat{D}_\xi{\hat{a}}^\dag \hat{D}_\xi^\dag={\hat{a}}^\dag-\xi^\ast
\end{equation}

\begin{equation}
\hat{R}_{\omega_r}\hat{a}\hat{R}_{\omega_r}^\dag=\hat{a}e^{-{i\omega}_rt}
\end{equation}

\begin{equation}
\hat{R}_{\omega_r}{\hat{a}}^\dag \hat{R}_{\omega_r}^\dag={\hat{a}}^\dag e^{{i\omega}_rt}
\end{equation}

The effective Hamiltonian now has the form $\hat{\mathcal{H}}_{\text{eff}}=\hat{U}\hat{\mathcal{H}}_{\text{tot}}\hat{U}^\dag+i(\partial_t\hat{U})\hat{U}^\dag$, where $\hat{\mathcal{H}}_{\text{tot}}=\hat{\mathcal{H}}_0+\hat{\mathcal{H}}_d$. Since the nonlinearities are weak, we determine the displacement $\xi$ from the linear part of the Hamiltonian alone. The transformed Hamiltonian can be written as:

\begin{equation}
\begin{aligned}
\hat{\mathcal{H}}_{\text{eff}}=&\;\omega_{\text{KO}}\left({\hat{a}}^\dag e^{i\omega_r t}-\xi^\ast\right)\left(\hat{a}e^{-i\omega_r t}-\xi\right)+\sum_{n\geq3}{g_n\left(\hat{a}e^{-i\omega_r t}-\xi+\text{h.c.}\right)^n}\\
&+2\cos{(\omega_d t)}\left(\epsilon_d\left({\hat{a}}^\dag e^{i\omega_r t}-\xi^\ast\right)+\epsilon_d^\ast\left(\hat{a}e^{-i\omega_r t}-\xi\right)\right)+i(\partial_t\hat{U})\hat{U}^\dag
\end{aligned}
\end{equation}

The contribution of the time-dependent unitary to the effective Hamiltonian is:  

\begin{equation}
    \begin{aligned}
    i(\partial_t\hat{U})\hat{U}^\dag&=i \left(i\omega_r \hat{a}^\dag\hat{a}+\hat{R}_{\omega_r}\left(\dot{\xi}{\hat{a}}^\dag-{\dot{\xi}}^\ast \hat{a}\right)\hat{R}_{\omega_r}^\dag\right) \\ &=-\omega_r {\hat{a}}^\dag\hat{a}+i\dot{\xi}{\hat{a}}^\dag e^{i\omega_r t}-i{\dot{\xi}}^\ast\hat{a}e^{-i\omega_r t}
    \end{aligned}
\end{equation}

The oscillator will be displaced until the coefficient of $\hat{a}^\dag$ is equal to zero

\begin{equation}
    \left(i\dot{\xi}+2\epsilon_d\cos(\omega_d t)-\omega_{\text{KO}}\xi\right)\hat{a}^\dag e^{i\omega_r t}=0
    \quad \Rightarrow \quad
    \xi=\frac{\epsilon_d}{\omega_{\text{KO}}+\omega_d}e^{i\omega_d t}+\frac{\epsilon_d}{\omega_{\text{KO}}-\omega_d}e^{-i\omega_d t}
\end{equation}

Inside the nonlinear terms we have $\xi+\xi^\ast=\xi_de^{-{i\omega}_dt}+c.c.$, so we obtain for the displacement inside the nonlinear terms:

\begin{equation}
    \xi_d=\frac{\epsilon_d}{\omega_{\text{KO}}-\omega_d}+\frac{\epsilon_d^\ast}{\omega_d+\omega_{\text{KO}}}=\frac{2\epsilon_d \omega_d}{\omega^2_{\text{KO}}-\omega^2_d}
    \label{eq:displacement_for_charge_driving}
\end{equation}

Because the charge drive couples through $\hat{n}$, $\epsilon_d$ is purely imaginary, giving $\epsilon_d^*=-\epsilon_d$. Choosing the frame frequency to be the KO frequency ($\omega_r=\omega_{\rm KO}$) makes the Stark shift equal to the final detuning. The effective Hamiltonian now simplifies in the displaced rotating frame to:

\begin{equation}
    \hat{\mathcal{H}}_{\text{eff}}=\delta{\hat{a}}^\dag\hat{a}+\sum_{n\geq3}{g_n\left(\hat{a}e^{-i\omega_r t}-\xi_de^{-i\omega_d t}+\text{h.c.}\right)^n}
    \label{eq:displaced_Hamiltonian}
\end{equation}

 \begin{figure}[t!]
    \includegraphics[width=0.99\textwidth]{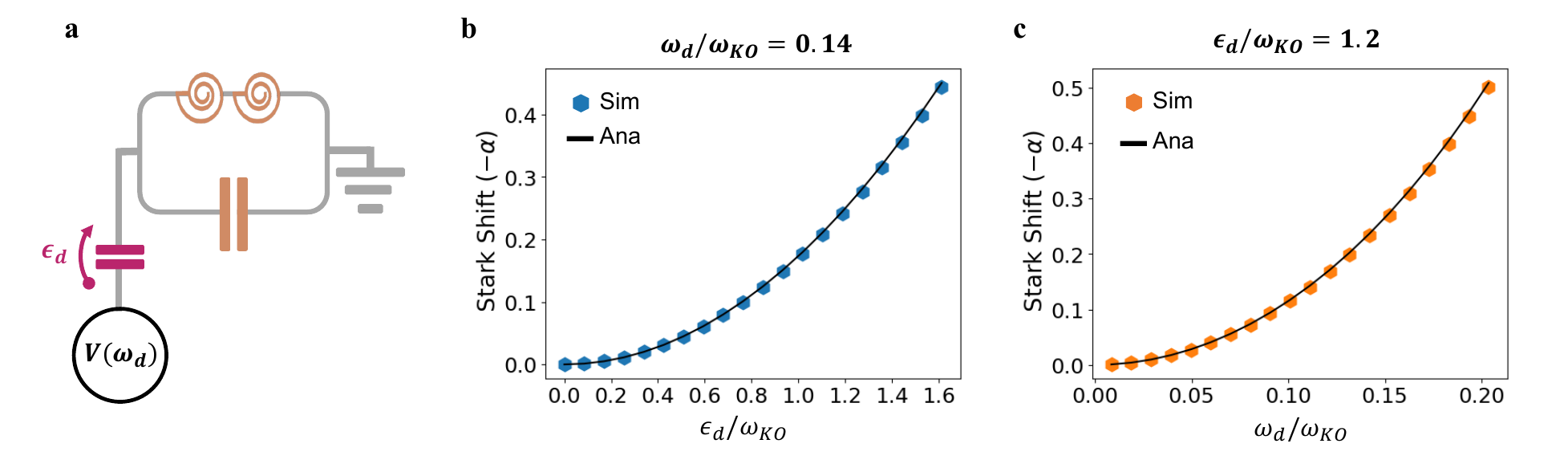}
    \caption{\textbf{Floquet simulation of the AC Stark shift.} \textbf{a,} An external voltage source modulates the gate charge on the double-SNAIL KO at a frequency $\omega_d$, which is modeled as the single-photon drive $\epsilon_d$. \textbf{b,} The AC Stark shift as a function of the drive amplitude $|\epsilon_d|$ at low frequency. The quadratic scaling results from two drive photons participating in this process. The simulation is based on the shift in the Floquet quasi-energies in the presence of the drive, and the analytical trend follows Eq.~\ref{eq:Stark_shift_Hamiltonian}. This treatment breaks down when the drive amplitude approaches the KO frequency with the maximum Stark shift much smaller than .\textbf{c,} The Stark shift as a function of the drive frequency $\omega_d$. The quadratic scaling is attributed to the near-cancellation of the rotating and counter-rotating terms in Eq.~\ref{eq:displacement_for_charge_driving}.}
    \label{fig:Floquet_simulation}
\end{figure}

Now, depending on the nonlinearity of the oscillator and the frequency of the drive, we can expand this expression and apply the RWA to obtain the Hamiltonian describing the wave mixing process. For the special case of the AC Stark shift, and using the formula for the anharmonicity derived in the previous Supplemental Note~\ref{note:snail_nonlinearity}, $\alpha=-12(g_4-5g_3^2/\omega_{\text{KO}})$, we get:

\begin{equation}
    \hat{\mathcal{H}}_s= -2 \alpha |\xi_d|^2  \hat{a}^\dagger \hat{a} \approx -8 \alpha \left(\frac{|\epsilon_d| \omega_d}{\omega^2_{\text{KO}}}  \right)^2 \hat{a}^\dagger \hat{a}
    \label{eq:Stark_shift_Hamiltonian}
 \end{equation}

 The results of this Hamiltonian with the numerical Floquet simulation are shown in Fig.~\ref{fig:Floquet_simulation}. The simulation is carried out using the Python package QuTiP \cite{qutip5}. The model uses the SNAIL potential up to the fourth-order nonlinearity and the charge drive in Eq.~\ref{eq:charge_drive} with ($\epsilon_d^*=-\epsilon_d$) without applying the RWA at any step. In parts \textbf{b,c}, the Stark shift builds up quadratically with the amplitude and frequency as explained in the main text. The quadratic scaling with the drive frequency traces back to the cancellation of the rotating and counter-rotating terms in Eq.~\ref{eq:displacement_for_charge_driving}. However, the unexpected result is the robustness of the oscillator to the drive strength exceeding the oscillator frequency to explain the Stark shift in the measured device. This robustness also enables the fast beam-splitter interactions between the KOs in the dual-rail encoding.
\newpage
\suppnote{Dressed states and KO readout}{note:dressed_states}

\noindent The dressed-state representation of the circuit modes enables us to understand the nonlinear interaction between the circuit modes, starting with the linear static coupling. This also clarifies the dispersive (i.e., cross-Kerr) interaction between the KO and the readout mode, as well as the beam-splitter interaction used to reset and cool the KO. In this section, we examine the Schrieffer--Wolff transformation that diagonalizes the full Hamiltonian via an appropriate unitary transformation. Starting with the KO--resonator Hamiltonian in the absence of microwave drives, we have:

\begin{equation}
    \hat{\mathcal{H}}_{\text{KO,R}}=\omega_R \hat{b}^\dagger \hat{b}+\omega_{\text{KO}} \hat{a}^\dagger \hat{a}-g_{\text{KO,R}}(\hat{a}^\dagger-\hat{a})(\hat{b}^\dagger-\hat{b})  +\sum{g_n\left(\hat{a}+\hat{a}^\dagger \right)^n}
    \label{eq:KO_resonator_Hamiltonian}
\end{equation}

Because the nonlinearity is weak, we seek a transformation that decouples the modes, which depend only on the coupling strength $g$ and the detuning $\Delta=\omega_R-\omega_{\text{KO}}$ within RWA. Our trial unitary is generated by the operator $A=\hat{a}^\dagger \hat{b}-\hat{a}\hat{b}^\dagger$ and transforms the Hamiltonian to the dressed basis according to:

\begin{equation}
    \hat{\mathcal{H}}_{\text{eff}}=e^{\lambda A} \hat{\mathcal{H}}_{\text{KO,R}} e^{-\lambda A}=\hat{\mathcal{H}}_{\text{KO,R}}+\lambda[A, \hat{\mathcal{H}}_{\text{KO,R}}] +\frac{\lambda^2}{2}[A,[A, \hat{\mathcal{H}}_{\text{KO,R}}]]+O\left(\lambda^3\right)
\end{equation}

The transformed annihilation operators, for example, are given by:

\begin{equation}
    e^{\lambda A} \hat{a} e^{-\lambda A}=\hat{a}+\lambda[A, \hat{a}]+\frac{\lambda^2}{2}[A,[A, \hat{a}]]+O\left(\lambda^3\right) = \cos (\lambda) \hat{a}-\sin (\lambda) \hat{b}
\end{equation}

\begin{equation}
    e^{\lambda A} \hat{b} e^{-\lambda A}=\hat{b}+\lambda[A, \hat{b}]+\frac{\lambda^2}{2}[A,[A, \hat{b}]]+O\left(\lambda^3\right) = \cos (\lambda) \hat{b}+\sin (\lambda) \hat{a}
\end{equation}

For simplicity, we will assume the detuning $\Delta$ to be much larger than the anharmonicity $\alpha$ to neglect the self-Kerr while diagonalizing the Hamiltonian. Applying the unitary explicitly we get: 

\begin{equation}
    \begin{aligned}
\hat{\mathcal{H}}_{\text{eff}} &\approx \omega_R  (\cos (\lambda) \hat{b}^\dagger+\sin (\lambda) \hat{a}^\dagger)( \cos (\lambda) \hat{b}+\sin (\lambda) \hat{a})\\& +
    \omega_{\text{KO}} (\cos (\lambda) \hat{a}^\dagger-\sin (\lambda) \hat{b}^\dagger)( \cos (\lambda) \hat{a}-\sin (\lambda) \hat{b})
    \\ & -g_{\text{KO,R}}(\cos (\lambda) \hat{a}^\dagger-\sin (\lambda) \hat{b}^\dagger) (\cos (\lambda) \hat{b}+\sin (\lambda) \hat{a})+\text{h.c.} \\ &  +\sum{g_n\left( \cos (\lambda) \hat{a}-\sin (\lambda) \hat{b}+\text{h.c.} \right)^n}
    \end{aligned}
\end{equation}

Now choosing $\lambda$ such that the linear coupling between the modes vanishes we get:

\begin{equation}
    \lambda=-\frac{1}{2} \arctan(2g_{\text{KO,R}}/\Delta)\approx -g_{\text{KO,R}}/\Delta
\end{equation}

So the effective Hamiltonian becomes (absorbing the Lamb shifts $\pm g^2_{\text{KO,R}}/\Delta$ into the new frequencies):

\begin{equation}
    \hat{\mathcal{H}}_{\text{eff}}\approx \omega_R \hat{b}^\dagger \hat{b}+\omega_{\text{KO}} \hat{a}^\dagger \hat{a} + \sum{g_n\left( \hat{a}+\frac{g_{\text{KO,R}}}{\Delta} \hat{b}+\text{h.c.} \right)^n} \approx \omega_R \hat{b}^\dagger \hat{b}+\omega_{\text{KO}} \hat{a}^\dagger \hat{a}-\frac{\alpha}{2} \hat{a}^{\dagger 2}\hat{a}^2 -2\alpha \left(\frac{g_{\text{KO,R}}}{\Delta}\right)^2 \hat{a}^\dagger  \hat{a} \hat{b}^\dagger\hat{b}
\end{equation}

\begin{figure}[t!]
    \includegraphics[width=0.99\textwidth]{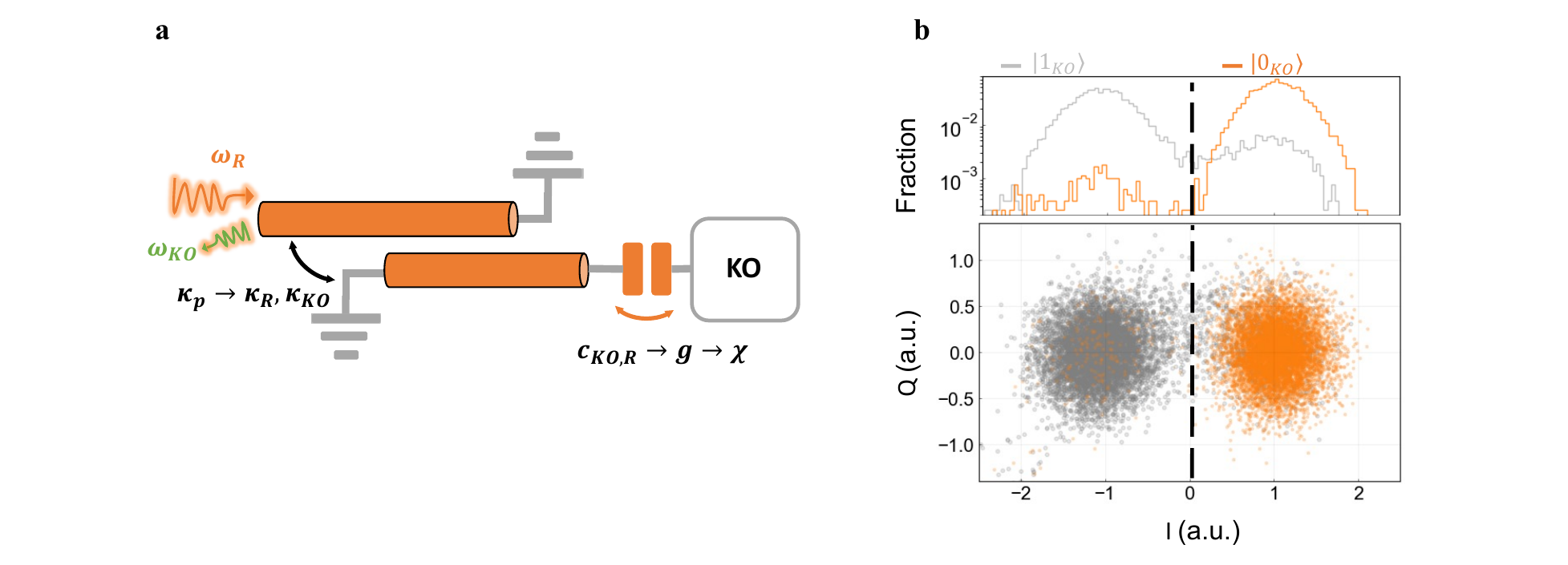}
    \caption{\textbf{KO readout.} \textbf{a,} The KO readout circuit consists of two coplanar waveguide (CPW) resonators: a readout mode and a Purcell filter. The capacitive coupling between the KO and the readout mode reduces to the dispersive interaction $\chi$. The decay rate of the Purcell filter $\kappa_P$ leads to the induced decay rates $\kappa_R$ in the readout mode and $\kappa_{\text{KO}}$ in the KO, which we seek to minimize. \textbf{b,} The reflected readout signal at the readout resonator frequency $\omega_R$ experiences a phase shift depending on the state of the KO. The data are rotated and displaced to center the discrimination line along the $ y$-axis. The top panel shows the histogram plot of the real part. Because we used a long readout time to maximize ground-state measurement fidelity, the excited state decays significantly to the ground state.}
    \label{fig:Readout}
\end{figure}

Using the RWA to simplify the expansion in the last step and neglecting the induced Kerr nonlinearity in the readout mode. The dispersive shift in the last term is $\chi=-2\alpha \left(\frac{g_{\text{KO,R}}}{\Delta}\right)^2$. This term can be interpreted as a frequency shift in the readout mode due to excitations in the KO. So, by probing the phase of the reflected signal at the readout resonator frequency $\omega_R$, we can infer the state of the KO. The schematic of this interaction is shown in Fig.~\ref{fig:Readout} \textbf{a}. In part \textbf{b} of the figure, we show an example of this readout signal used to infer the thermal population of the KO. To maximize ground-state measurement fidelity, we perform the readout for $7 \, \mu$s, which causes excess decay of the excited state. The steady-state thermal population of the KO from this measurement is estimated to be $n_{\text{th}}=2.5 \%$. The active reset presented in the main text reduces the thermal population of the KO to $0.7 \%$. This value is typical due to measurement-induced state transitions (MIST) \cite{MIST_PhysRevApplied,MIST_prx,MIST_PhysRevLett} and remains substantially larger than the ideal classification infidelity of $0.01 \%$ using \cite{krantz2019quantum}:

\begin{equation}
    \epsilon_{\text{class}}=\frac{1}{2} \text{erfc}\left(\frac{\text{SNR}}{\sqrt{8}}\right)
\end{equation}

With the signal-to-noise ratio:

\begin{equation}
    \text{SNR}=\frac{|\langle S_1 \rangle -\langle S_0\rangle|}{|\sigma_0|}
\end{equation}

with $\langle S_i \rangle$ being the mean of the data for a certain outcome and $\sigma^2_0$ being the variance of the distribution for the ground state. This high separation, despite the small anharmonicity, was enabled by the use of a superconducting quantum-limited amplifier \cite{macklin2015twpa} and a narrow readout resonator (i.e., small $\kappa_R$).

The dedicated Purcell filters in the readout circuit are used to suppress the Purcell decay of the qubit relative to the readout mode. This can be modeled using the linearized, non-Hermitian effective KO Hamiltonian in the dispersive regime, as follows ($\hbar=1$):

\begin{equation}
    \begin{split}
    &\hat{\mathcal{H}}_0\approx \omega_P{\hat{p}}^\dag\hat{p}+\omega_R{\hat{b}}^\dag\hat{b}+\omega_{\text{KO}}{\hat{a}}^\dag\hat{a}\\
    &\hat{\mathcal{H}}_c\approx g_{\text{KO,R}}\left({\hat{a}}^\dag\hat{b}+\hat{a}{\hat{b}}^\dag\right)+g_{\text{P,R}}\left({\hat{b}}^\dag\hat{p}+\hat{b}{\hat{p}}^\dag\right)\\
    &\hat{\mathcal{H}}_\kappa\approx -\frac{i\kappa}{2}{\hat{p}}^\dag\hat{p}
    \end{split}
\end{equation}

where $\kappa$ is the decay rate of the Purcell filter in the mode $\hat{p}$. We expect the coupling and detuning between the modes to determine the hybridization, since the Purcell filter is realized in the good-cavity limit. To simplify the diagonalization of the system, we can move to a rotating frame at the frequency of the readout resonator and focus on the single-excitation subspace:

\begin{equation}
    \begin{pmatrix}
\Delta_{\text{KO,R}} & g_{\text{KO,R}} & 0 \\
g_{\text{KO,R}} & 0 & g_{\text{P,R}} \\
0 & g_{\text{P,R}} & \frac{-i\kappa}{2} +\Delta_{\text{P,R}}
\end{pmatrix}
\rightarrow
\begin{vmatrix}
\Delta_{\text{KO,R}} - \lambda & g_{\text{KO,R}} & 0 \\
g_{\text{KO,R}} & -\lambda & g_{\text{P,R}} \\
0 & g_{\text{P,R}} & \frac{-i\kappa}{2} + \Delta_{\text{P,R}} - \lambda
\end{vmatrix} = 0
\end{equation}

\begin{equation}
    -g_{\text{KO,R}}^2\left(\frac{-i\kappa}{2}+\Delta_{\text{P,R}}-\lambda\right)-\lambda\left(\Delta_{\text{KO,R}}-\lambda\right)\left(\frac{-i\kappa}{2}+\Delta_{\text{P,R}}-\lambda\right)-g_{\text{P,R}}^2\left(\Delta_{\text{KO,R}}-\lambda\right)=0
\end{equation}

To estimate the decay rate of the qubit, we can first factor out the leading correction that comes from the qubit-resonator coupling and then expand for the correction $\delta$ to get the imaginary part of the eigenvalue, which gives the effective decay rate:

\begin{equation}
\lambda=\Delta_{\text{KO,R}}+\frac{g_{\text{KO,R}}^2}{\Delta_{\text{KO,R}}}+\delta=\lambda_0+\delta
\end{equation}

Now expanding and defining $\Delta_{\text{P,Q}}=\Delta_{\text{P,R}}-\Delta_{\text{KO,R}}$, we get: 

\begin{equation}
    \begin{split}
    &-g_{\text{KO,R}}^2\left(\frac{-i\kappa}{2}+\Delta_{\text{P,Q}}-\frac{g_{\text{KO,R}}^2}{\Delta_{\text{KO,R}}}-\delta\right)\\&+\left(\frac{g_{\text{KO,R}}^2}{\Delta_{\text{KO,R}}}+\delta\right)\left(\delta+\frac{g_{\text{KO,R}}^2}{\Delta_{\text{KO,R}}}+\Delta_{\text{KO,R}}\right)\left(\frac{-i\kappa}{2}+\Delta_{\text{P,Q}}-\frac{g_{\text{KO,R}}^2}{\Delta_{\text{KO,R}}}-\delta\right)\\&+g_{\text{P,R}}^2\delta+g_{\text{P,R}}^2\frac{g_{\text{KO,R}}^2}{\Delta_{\text{KO,R}}}=0
    \end{split}
\end{equation}

Now we can write $\delta= \delta_n/\delta_d$ and simplify each term to get for the numerator:

\begin{equation}
    \begin{split}
    \delta_n&=-g_{\text{P,R}}^2\frac{g_{\text{KO,R}}^2}{\Delta_{\text{KO,R}}}+g_{\text{KO,R}}^2\left(\frac{-i\kappa}{2}+\Delta_{\text{P,Q}}-\frac{g_{\text{KO,R}}^2}{\Delta_{\text{KO,R}}}\right)\\&-g_{\text{KO,R}}^2\left(\frac{g_{\text{KO,R}}^2}{\Delta_{\text{KO,R}}^2}+1\right)\left(\frac{-i\kappa}{2}+\Delta_{\text{P,Q}}-\frac{g_{\text{KO,R}}^2}{\Delta_{\text{KO,R}}}\right)\\&  =-g_{\text{P,R}}^2\frac{g_{\text{KO,R}}^2}{\Delta_{\text{KO,R}}}-g_{\text{KO,R}}^2\frac{g_{\text{KO,R}}^2}{\Delta_{\text{KO,R}}^2}\left(\frac{-i\kappa}{2}+\Delta_{\text{P,Q}}-\frac{g_{\text{KO,R}}^2}{\Delta_{\text{KO,R}}}\right)
    \end{split}
\end{equation}

The denominator simplifies to:
\begin{equation}
    \begin{split}
\delta_d&= g_{\text{KO,R}}^2+g_{\text{P,R}}^2+\left(\frac{g_{\text{KO,R}}^2}{\Delta_{\text{KO,R}}}+\Delta_{\text{KO,R}}\right)\left(\frac{-i\kappa}{2}+\Delta_{\text{P,Q}}-\frac{g_{\text{KO,R}}^2}{\Delta_{\text{KO,R}}}\right)\\&+\left(\frac{g_{\text{KO,R}}^2}{\Delta_{\text{KO,R}}}\right)\left(\frac{-i\kappa}{2}+\Delta_{\text{P,Q}}-\frac{g_{\text{KO,R}}^2}{\Delta_{\text{KO,R}}} \right)-\left(\frac{g_{\text{KO,R}}^2}{\Delta_{\text{KO,R}}}\right)\left(\frac{g_{\text{KO,R}}^2}{\Delta_{\text{KO,R}}}+\Delta_{\text{KO,R}}\right) \\& \approx\Delta_{\text{KO,R}}\left(\frac{-i\kappa}{2}+\Delta_{\text{P,Q}}-\frac{g_{\text{KO,R}}^2}{\Delta_{\text{KO,R}}}\right)
\end{split}
\end{equation}

Finally, we can evaluate the imaginary part of the eigenvalue:

\begin{equation}
 \text{Im}{\left(\frac{\delta_n}{\delta_d}\right)} \approx - \frac{g_{\text{P,R}}^2 g_{\text{KO,R}}^2}{\Delta_{\text{KO,R}}^2 \Delta_{\text{P,Q}}^2} \frac{\kappa}{2} 
\end{equation}

Or equivalently:

\begin{equation}
    \begin{split}
    &\kappa_R\approx\kappa\frac{g_{\text{P,R}}^2}{\Delta_{\text{P,R}}^2}\\
    & \kappa_Q\approx\kappa\frac{g_{\text{P,R}}^2}{\Delta_{\text{P,Q}}^2}\frac{g_{\text{KO,R}}^2}{\Delta_{\text{KO,R}}^2}
    \end{split}
\end{equation}

This suppresses the standard Purcell limit by an extra factor of $\Delta_{\text{P,R}}^2 / \Delta_{\text{P,Q}}^2 \le 10^{-2}$ which leads to a net Purcell limit of tens of milliseconds which is much higher than the tens of microseconds lifetime of the excited states.

\newpage
\suppnote{Beam-splitter interaction and noise analysis in the dual-rail encoding}{note:dual_rail_noise_analysis}

\noindent In the dual-rail encoding scheme, a strong beam-splitter drive enables us to perform logical operations while simultaneously suppressing both the action of single-photon loss inside the computational subspace and the KOs' frequency fluctuation. The latter is true for both low-frequency noise and high-frequency noise as long as the drive amplitude squared $\epsilon^2_{\text{BS}}$ is sufficiently larger than the variance of the noise $\langle \delta^2 \rangle$. In this note, we first explain the origin of this beam-splitter drive and the phase alignment when simultaneously driving both KOs. Next, we explain the suppression of single-photon loss inside the computational subspace. Finally, we examine the suppression of the KOs' frequency fluctuations due to low-frequency and high-frequency noise in the strong-driving regime.

Following Supplementary Notes \ref{note:device},\ref{note:dressed_states} , and \ref{note:wave_mixing}, we can represent the Hamiltonian of the coupled KOs in the dressed basis and then apply the appropriate unitary transformations to absorb the low-frequency drives as displacements that mediate the wave mixing process (neglecting crosstalk, as explained in supplementary note 1). To start, we consider two coupled KOs at a rate $g_{0,1}\approx -\frac{c_c}{2c_{\text{KO}}} \sqrt{\omega_0 \omega_1}$ and driven at frequency $\omega_d$:

\begin{equation}
    \begin{aligned}
    \hat{\mathcal{H}}_{\text{KOs}} &= \omega_0 \hat{a}_0^\dag \hat{a}_0 + \omega_1 \hat{a}_1^\dag \hat{a}_1+g_{0,1} (\hat{a}_0^\dagger-\hat{a}_0)(\hat{a}_1^\dagger-\hat{a}_1) \\
    & + 2i \epsilon_0 \cos{(\omega_d t)}   (\hat{a}_0^\dagger-\hat{a}_0) + 2i\epsilon_1 \cos{(\omega_d t)}  (\hat{a}_1^\dagger-\hat{a}_1) \\
    & + \sum_{n\geq3}{g_{n,0}\left(\hat{a}_0 +\text{h.c.}\right)^n} + \sum_{n\geq3}{g_{n,1}\left(\hat{a}_1 +\text{h.c.}\right)^n}
    \end{aligned}
\end{equation}

Decoupling the modes using a similar Schrieffer--Wolff transformation as in Note \ref{note:dressed_states} (i.e., $\lambda \approx \pm \frac{g_{0,1}}{\Delta_{\text{KO}}}$) and then applying the unitary transformations to absorb the drives as displacements in the nonlinear terms (similar to Note \ref{note:wave_mixing}), we get:

\begin{equation}
    \begin{aligned}
    \hat{\mathcal{H}}_{\text{eff}}& \approx \delta_0 {\hat{a}}_0^\dag{\hat{a}}_0+\delta_1 {\hat{a}}_1^\dag{\hat{a}}_1 \\ 
    & +\sum_{n\geq3}{g_{n,0}\left(\hat{a}_0 e^{-i\omega_{r,0} t}-\xi_0 e^{-i\omega_d t}-\frac{g_{0,1}}{\Delta_{\text{KO}}} \hat{a}_1 e^{-i\omega_{r,1} t}+\text{h.c.}\right)^n} \\
    &+ \sum_{n\geq3}{g_{n,1}\left(\hat{a}_1 e^{-i\omega_{r,1} t}-\xi_1 e^{-i\omega_d t}+\frac{g_{0,1}}{\Delta_{\text{KO}}} \hat{a}_0 e^{-i\omega_{r,0} t}+\text{h.c.}\right)^n} 
    \end{aligned}
\end{equation}

In the absence of the drives, the fourth-order nonlinearity will give rise to the cross-Kerr interaction $\chi_{0,1}$ between the KOs. However, in the presence of the drive, the third-order nonlinearity gives rise to the beam-splitter interaction $\epsilon_{\rm BS}$ between the KOs, and the fourth-order nonlinearity adds the Stark shift. The beam-splitter drive can be obtained by setting the frame frequencies to the KOs frequencies (i.e., $\omega_{r,i} = \omega_i$) and the drive frequency to match the KOs frequency difference ($\omega_d = \omega_0 - \omega_1$). As a result, the interaction between the KOs within the RWA is given by:

\begin{equation}
    \hat{\mathcal{H}}_{\text{int}} \approx \epsilon_{\text{BS}} \hat{a}_0^\dagger \hat{a}_1 + \epsilon^{*}_{\text{BS}} \hat{a}_0 \hat{a}_1^\dagger + \chi_{0,1} \hat{a}_0^\dagger \hat{a}_0 \hat{a}_1^\dagger \hat{a}_1 
\end{equation}

\begin{equation}
    \epsilon_{\text{BS}}= 6g_{3,0}\xi_0 \frac{g_{0,1}}{\Delta_{\text{KO}}}-6g_{3,1}\xi_1 \frac{g_{0,1}}{\Delta_{\text{KO}}}
\end{equation}

\begin{equation}
    \chi_{0,1} = -2 (\alpha_0 + \alpha_1) \left(\frac{g_{0,1}}{\Delta_{\text{KO}}}\right)^2
\end{equation}

This shows that we need to drive the KOs out of phase (i.e., opposite displacements) so the beam-splitter contributions add up (due to the opposite hybridizations). In practice, small phase differences due to wiring-length differences to the KOs can be accounted for in control-signal calibration \cite{Qubic}. However, this requires microwave crosstalk between the KOs and the drive lines to be small enough to address a single KO with each line. The cross-Kerr part of the interaction plays no role in the dual-rail encoding since it has zero projection onto the computational subspace.

Now we consider the effect of two error channels in the presence of the beam-splitter drive and the condition for their suppression. First, the action of single-photon loss inside the computational subspace can be analyzed using the effective non-Hermitian Hamiltonian (since quantum jumps map the code words to the ground state) to get: 

\begin{equation}
    \hat{\mathcal{H}}_{\text{pl}} = \epsilon_{\text{BS}} \hat{a}_0^\dagger \hat{a}_1 + \epsilon^{*}_{\text{BS}} \hat{a}_0 \hat{a}_1^\dagger -i\frac{ \Gamma_0}{2} \hat{a}_0^\dagger \hat{a}_0  -i \frac{\Gamma_1}{2} \hat{a}_1^\dagger \hat{a}_1 
\end{equation}

Now projecting this onto the computational subspace (assuming $\epsilon_{\text{BS}}$ to be real for simplicity):

\begin{equation}
    \hat{P}_{\rm DR}\hat{\mathcal{H}}_\text{\rm pl}\hat{P}_{\rm DR}=\epsilon_{\text{BS}}\hat{X}-i\frac{\Gamma_{\rm DL}}{2}\hat{I} -i\frac{\delta \Gamma}{4} \hat{Z}
\end{equation}

With $\delta \Gamma = \Gamma_0 - \Gamma_1, \Gamma_{\rm DL} = (\Gamma_0+\Gamma_1)/2$. The term proportional to the identity is the uniform leakage (erasure) rate. The other component of the loss couples to a non-commuting operator with the drive and hence they tend to suppress each other. Quantitatively, we can examine the eigenvalues of the effective Hamiltonian (neglecting leakage):

\begin{equation}
    E_\pm = \pm \sqrt{\epsilon_{\text{BS}}^2- \delta \Gamma^2/16} 
\end{equation}

The eigenvalues are real when $4|\epsilon_{\rm BS}|>|\delta\Gamma|$. In that regime the two KOs are mixed faster than they can be distinguished by decay, so $\delta\Gamma$ renormalizes the Rabi frequency rather than damping the Bloch vector inside the code space. The identity term proportional to $\Gamma_{\rm DL}$ remains as the erasure rate. This is the PT-symmetric regime discussed in Ref.~\cite{Naghiloo2019NaturePhysics}.

Similarly, we can analyze the effect of frequency fluctuations by noting that the drive acts along an axis perpendicular to the noise. Considering a generic time-dependent detuning $\delta(t)$, we get:

\begin{equation}
    \hat{\mathcal{H}}_{\text{det}}=\epsilon_{\rm BS}\hat{X}+\delta(t) \hat{Z}
\end{equation}

With $\delta(t) = [\delta_0 (t) - \delta_1 (t)]/2$. There are three limits to this equation of interest in the strong-driving regime (i.e., $\epsilon^2_{\text{BS}} \gg |\langle \delta^2(t) \rangle|$). First, when the detuning is changing at a rate equal to the splitting $2 \epsilon_{\text{BS}}$, this flips the pseudospin state and acts as a bit-flip error when it is incoherent \cite{Decoherence_Superconducting_Qubit_PhysRevB}. In practice, because the noise is typically $1/f$, this effect is suppressed as the drive strength increases. Second, when the fluctuation changes much more slowly than a single experimental run (i.e., quasistatic), it acts as a dephasing error. To see this, we first look at the accumulated phase $\delta \varphi(t)$ due to the energy shift $\delta E =\delta^2_{\text{qs}}/\epsilon_{\text{BS}} \rightarrow \delta \varphi(t)=\delta E \, t$. The evolution of the logical observable $\hat{Z}$ is proportional to $\Re[e^{-i 2\epsilon_{\text{BS}}t} \langle e^{-i \delta \varphi(t)} \rangle]$. Now because $\delta_{\text{qs}}$ is assumed to be normally distributed with variance $\langle \delta^2_{\text{qs}} \rangle$, the decay of the logical observable follows:

\begin{equation}
    \langle e^{-i \delta \varphi(t)} \rangle \approx \langle e^{-i \frac{\delta^2_{\text{qs}}}{\epsilon_{\text{BS}}}t} \rangle  = \frac{1}{\left[1+i \frac{2\langle \delta^2_{\text{qs}} \rangle}{ \epsilon_{\text{BS}} }t\right]^{1/2}} =\frac{1}{\left[1+ (\Gamma_{\phi}t)^2 \right]^{1/4} } e^{-i\gamma_{\phi}(t)} 
\end{equation}

With $\Gamma_{\phi} = \frac{2\langle \delta^2_{\text{qs}} \rangle}{\epsilon_{\text{BS}}}$ and $\gamma_{\phi}(t) =\frac{1}{2} \arctan{(\Gamma_{\phi}t)}$. So, by increasing the drive strength, we can suppress the effect of quasistatic noise so it plays a negligible role in the error budget, because the power-law decay is much slower than the exponential decay associated with photon loss or high-frequency noise. 

Finally, we analyze the effect of Gaussian noise, which is still slow compared to the splitting $2\epsilon_{\text{BS}}$ but varies within a single experimental run with a single correlation time $\tau_c$ much smaller than any experimental timescales, starting with the autocorrelation function:

\begin{equation}
    \langle \delta(t) \delta(t+\tau) \rangle = \sigma^2 e^{-|\tau|/\tau_c}
\end{equation}

As long as the frequency fluctuations are small (i.e., $\sigma^2 \ll \epsilon^2_{\text{BS}}$) then using the second-order result with the zero-mean frequency fluctuation, we get:

\begin{equation}
    \delta\omega(t) = \frac{\delta(t)^2 -\sigma^2}{\epsilon_{\rm BS}}
\end{equation}

leads to the random phase error:

\begin{equation}
    \delta\varphi(t) = \int_0^t \delta\omega(t') dt'
\end{equation}

we can derive the decay using the Gaussian approximation:

\begin{equation}
    \langle e^{-i \delta\varphi(t)} \rangle= e^{- \frac{1}{2} \langle \delta\varphi(t)^2 \rangle}= \exp\left( - \frac{1}{2} \int_0^t \int_0^t \langle \delta\omega(t_1) \delta\omega(t_2) \rangle dt_1 dt_2 \right)
\end{equation}

Defining the autocorrelation function for the frequency fluctuation:

\begin{equation}
    K(\tau) = \langle \delta\omega(t) \delta\omega(t+\tau) \rangle = \frac{1}{\epsilon_{\rm BS}^2} \langle (\delta(t)^2 - \sigma^2)(\delta(t+\tau)^2 - \sigma^2) \rangle
\end{equation}

The higher-order Gaussian moments can be calculated according to:

\begin{equation}
    \langle \delta(t)^2 \delta(t+\tau)^2 \rangle = \langle \delta(t)^2 \rangle \langle \delta(t+\tau)^2 \rangle + 2\langle \delta(t)\delta(t+\tau) \rangle^2=\sigma^4 + 2\sigma^4 e^{-2|\tau|/\tau_c}
\end{equation}

Simplifying the autocorrelation function, we get:

\begin{equation}
    K(\tau) = \frac{1}{\epsilon_{\rm BS}^2} \left( 2\sigma^4 e^{-2|\tau|/\tau_c} \right) = \frac{2\sigma^4}{\epsilon_{\rm BS}^2} e^{-2|\tau|/\tau_c}
\end{equation}

Performing the double integral for time scales much longer than the correlation time, we get:

\begin{equation}
    \int_0^t \int_0^t K(t_1 - t_2) dt_1 dt_2 \approx 2t \int_0^\infty K(\tau) d\tau = \frac{2\sigma^4 \tau_c}{\epsilon_{\rm BS}^2} t
\end{equation}

which finally leads to the quadratically suppressed exponential decay with the rate:

\begin{equation}
    \Gamma_{\text{GN}} = \frac{\sigma^4 \tau_c}{\epsilon_{\rm BS}^2}
\end{equation}

In this work we focus on the quasistatic limit for simplicity but we expect the exact nature of the noise to be more complicated with contributions from high-frequency fluctuators.

\suppbibliography{manuscript_samp}

\end{document}